\documentclass[twocolumn,showpacs,preprintnumbers,amsmath,amssymb]{revtex4}
\usepackage{graphicx}
\usepackage{dcolumn}
\usepackage{bm}

\def\pj{\hspace{-.26cm}}

\def\thalf{{\textstyle{\frac{1}{2}}}}

\def\tquar{{\textstyle{\frac{1}{4}}}}

\def\ttquar{{\textstyle{\frac{3}{4}}}}

\def\s{\sigma}
\def\so{\sigma_0}
\def\pv{\vmg{\pi}}

\def\psv{\vmg{\psi}}
\def\Bmunu{\vmg{B_{\mu\nu}}}
\def\b{\vmg{b}}
\newcommand{\vm}[1]{\mbox{\bf#1}}
\newcommand{\vmg}[1]{\mbox{\boldmath$#1$}}

\newcommand{\bq}    {\begin{equation}}
\newcommand{\eq}    {\end{equation}}
\newcommand{\bqr} {\begin{eqnarray}}
\newcommand{\eqr} {\end{eqnarray}}

\begin{document}

\title{A chiral lagrangian with Broken Scale: \\
testing the restoration of symmetries in astrophysics and in the laboratory}

\author{Luca Bonanno} 
\author{Alessandro Drago}
\affiliation{Dipartimento di Fisica, Universit\`a di Ferrara and \\
INFN, Sezione di Ferrara, via Saragat 1, 44100 Ferrara, Italy}

\begin{abstract}
We study matter at high density and temperature using a chiral
lagrangian in which the breaking of scale invariance is regulated by
the value of a scalar field, called dilaton
\cite{Heide:1993yz,Carter:1995zi,Carter:1996rf,Carter:1997fn}.  We
provide a phase diagram describing the restoration of chiral and scale
symmetries. We show that chiral symmetry is restored at large
temperatures, but at low temperatures it remains broken at all
densities. We also show that scale invariance is more easily restored
at low rather than large baryon densities. The masses of vector mesons
scale with the value of the dilaton and their values initially
slightly decrease with the density but then they increase again for
densities larger than $\sim 3 \, \rho_0$.  The pion mass increases
continuously with the density and at $\rho_0$ and T=0 its value is
$\sim$ 30 MeV larger than in the vacuum.  We show that the model is
compatible with the bounds stemming from astrophysics, as e.g. the one
associated with the maximum mass of a neutron star. The most striking
feature of the model is a very significant softening at large
densities, which manifests also as a strong reduction of the adiabatic
index. While the softening has probably no consequence for Supernova
explosion via the direct mechanism, it could modify the signal in
gravitational waves associated with the merging of two neutron stars.
\end{abstract}


\maketitle

\section{Introduction}

A very hot problem in hadronic and nuclear physics is to investigate
the Equation of State (EOS) of matter at large density and/or temperature.
Several experiments have been proposed and will likely be performed in the
next years. One experiment has been planned at facility FAIR at GSI 
\cite{Senger:2004jw}, other experiments could be performed at RHIC (Brookhaven) and
at the Nuclatron in Dubna. In all these experiments Heavy Ion Collisions 
(HICs) will take place
at energies of the order of a few ten A GeV, which correspond to the regime
at which maximum baryon densities can be reached. 
Another important source of information concerning matter at high densities
is the study of the astrophysics of compact stars, where the temperatures are moderate
(typically lower than a few ten MeV), but the baryon density can reach several times
nuclear matter saturation density $\rho_0$.
It is therefore extremely important to investigate the EOS by using various
theoretical techniques, providing hints about the main features of matter at 
those extreme conditions. In a sense one is providing a map of the 
``new'' regions which experiments will likely explore.

In this work we use an effective model for
nuclear physics, as a tool for exploring theoretically the unknown regions.
The main feature of this model is that it is based on a chiral invariant lagrangian.
It is well known that it is not trivial to
introduce chiral symmetry in an effective nuclear lagrangian,
so that saturation properties of nuclei are well reproduced
together with the possibility to restore chiral symmetry at
large density and/or temperature.
For instance, the attempt of using the linear sigma model
to describe nuclear dynamics fails due to the impossibility of
reproducing the basic properties of nuclei \cite{Furnstahl:1995zb}.
The coefficients of the self-couplings of the scalar field are dictated by the form of the
potential and chiral symmetry restoration takes place already at 
about $\rho_0$.
More sophisticated approaches have been proposed in the literature.
A few works are based on a SU(2) chiral symmetry scheme,
either adopting the parity-doublet formulation 
\cite{Dexheimer:2007tn,Dexheimer:2008aj,Dexheimer:2008cv}
or introducing the scale invariance 
\cite{Mishustin:1993ub,Furnstahl:1995by}. Other works  
extend the symmetry to the strange sector 
\cite{Papazoglou:1997uw,Papazoglou:1998vr,Wang:2003cn,Dexheimer:2007df,Dexheimer:2008ax}.
Here we will explore the model introduced by the nuclear physics group
of the University of Minnesota
\cite{Heide:1993yz,Carter:1995zi,Carter:1996rf,Carter:1997fn}.
In that model chiral fields are present together with 
a dilaton field 
which reproduces the breaking of scale symmetry in QCD, but at a mean
field level.
In their first paper \cite{Heide:1993yz} they have shown that it is indeed
possible to reproduce the main features of closed shell nuclei; in a second
paper \cite{Carter:1995zi} the phenomenology of chiral fields, including
their interaction with the nuclei, is well described; in a third paper 
\cite{Carter:1996rf} they discuss 
mesons at finite temperature showing in particular that chiral
symmetry is restored at large temperature; finally in
Ref.~\cite{Carter:1997fn} they discuss symmetric
nuclear matter at finite density and finite temperature.

In our paper we extend the previous 
analysis providing rather detailed previsions concerning
various quantities which can be measured in the laboratory, like
e.g. the compressibility of matter and the behavior of the masses
of meson fields at large densities. We will study also the 
dependence of those quantities on the isospin fraction.
Moreover, we will test this chiral model
at large baryon and isospin densities which are relevant for studying various
astrophysical problems, as the structure and formation of compact stars.
In this way we complete the analysis provided in Ref.\cite{Bonanno:2007kh} 
where we compared the compressibility computed in various nuclear and hadronic models
for the EOS of matter.

A few caveats concerning the validity of our results at very large densities
and temperatures are in order:
clearly hyperonic and/or quark degrees of freedom should appear
in the high energy density regime. On the other hand
before investigating even more complicated lagrangians it is worth 
analyzing in detail the outcome of a purely hadronic model.
It is also important to remark that the restoration of scale invariance
was analyzed only in Refs.~\cite{Carter:1996rf,Carter:1997fn} obtaining a unrealistically
large critical temperature. In our paper we show that 
the inclusion of thermal fluctuations of the vector meson fields reduces
the critical temperature for scale restoration to a more realistic value and
provides a rather new and interesting phase diagram. 
Hyperons are not included in the present work in order to keep the
discussion of chiral symmetry and scale invariance restoration as simple
as possible. A more realistic model should incorporate also hyperons,
following e.g. the schemes proposed in 
Refs.~\cite{Papazoglou:1997uw,Papazoglou:1998vr,Wang:2003cn,Dexheimer:2007df,Dexheimer:2008ax}.
Notice finally that the lagrangian we are investigating mimics the behavior 
of the gluon condensate through the dynamics of the dilaton field. 
Since the value of the gluon condensate can determine if 
the system settles in the hadronic or in the quark-gluon phase \cite{Drago:2001gd}, this model 
is particularly suitable to be extended to the quark-gluon
sector in order to investigate
deconfinement, what will be done in a future work.

The structure of the paper is the following.
In Sec.~\ref{model} we will describe the model we are using,
in Sec.~\ref{lab} 
we discuss the EOS as it can be tested in lab experiments,
in Sec.~\ref{astro}
we show the implications for astrophysics. 
Finally, in Sec.~\ref{Conclusions} we will present our conclusions.

\section{The chiral-dilaton model}
\label{model}
\subsection{Scale invariance}

The aim of many effective lagrangians for nuclear physics
is to mimic the behavior of QCD in the strong regime, where chiral
and scale symmetries are broken. While the breaking of chiral symmetry 
is connected with the generation of fermionic masses, the breaking
of scale invariance is due to quantum corrections 
(the so-called scale or trace anomaly)
and is likely associated
with the formation of a gluon condensate. More explicitly,
the violation of scale invariance in QCD 
corresponds to a non-conservation of the dilatation
current, which reads:
\bq
\langle \partial_\mu j^\mu_{QCD}\rangle=\theta^{\mu}_{\mu}
= \frac{\beta(g_s)}{2g_s}F^a_{\mu\nu}(x)F^{a\mu\nu}(x) \, ,\label{scale}
\eq
where $j^\mu_{QCD}$ is the dilatation current in QCD, $\theta_\mu^\nu$ is
the ``improved'' energy-momentum tensor, 
$F^a_{\mu\nu}(x)$ is the gluon field strength tensor and $\beta(g_s)$ is
the QCD beta function.

In the approach of Schechter, Migdal and Shifman \cite{Schechter:1980ak,Migdal:1982jp}
a scalar field representing
the gluon condensate is introduced and its dynamics is regulated by a potential 
chosen so that it reproduces (at mean field level) the divergence of the scale current 
which in QCD is given by eq.~(\ref{scale}).
The potential of the dilaton field is therefore determined by the equation:
\begin{equation}
\theta^{\mu}_{\mu}=4V(\phi)-\phi\frac{\partial V}{\partial\phi}
=4\epsilon_{\rm vac}\left(\frac{\phi}{\phi_0}\right)^{\!4}\, \label{schechter},
\end{equation}
where the parameter $\epsilon_{\rm vac}$ represents the vacuum energy.

The previous approach holds in the case of pure gauge QCD
where, at first loop, the violation of the scale invariance reads:
\bq
\theta^{\mu}_{\mu} = - \frac{11 N_c}{96 \pi^2}\langle {g_s}^2 F^2\rangle \, ,
\eq
where $N_c$ is the number of colors.
To take into account
massless quarks a generalization was proposed in \cite{Heide:1992tk}, so that 
also chiral fields contribute to the trace anomaly. In this way the single scalar field
of eq.~(\ref{schechter}) is replaced by a set of scalar fields 
$\{\sigma,\pv,\phi\}$. The relative weight of the gluons and of the quarks
to the violation of scale invariance can be read from the QCD $\beta$-function, which
at first loop is:
\bq
\beta(g_s)=-\frac{11 {g_s}^3}{16 \pi^2}(1-\frac{2 n_f}{33})\, , \label{beta}
\eq
where $n_f$ is the number of massless flavors.

\subsection{The lagrangian}
\label{lagrangian}
The lagrangian of the chiral-dilaton model (CDM) reads:

\begin{eqnarray}
{\cal L}&=&\thalf\partial_{\mu}\sigma\partial^{\mu}
\sigma+\thalf\partial_{\mu}\vmg{\pi}\cdot\partial^{\mu}\vmg{\pi}
+\thalf\partial_{\mu}\phi\partial^{\mu}\phi
-\tquar\omega_{\mu\nu}\omega^{\mu\nu}\nonumber\\
&-&\tquar\Bmunu\cdot\Bmunu
+\thalf G_{\omega\phi}\phi^2 \omega_\mu\omega^\mu 
+\thalf G_{b\phi}\phi^2 \b_\mu\cdot\b^\mu\nonumber\\ 
&+&[(G_4)^2\omega_\mu\omega^\mu]^2-{\cal V}\\
&+&\bar{N}\left[\gamma^\mu(i\partial_{\mu}-g_\omega\omega_\mu
-\thalf g_{\rho}\b_{\mu}\cdot\vmg{\tau})
-g\sqrt{\s^2+\pv^2} \right] N \label{lb}\nonumber
\end{eqnarray}
where the potential governing the chiral and the dilaton field reads:
\begin{eqnarray}
{\cal V}&=& B\phi^4
\left(\ln\frac{\phi}{\phi_0}-\frac{1}{4}\right)
\hspace{-.73mm}-\hspace{-.73mm}\thalf B\delta\phi^4
\ln\frac{\sigma^2+\vmg{\pi}^2}{\sigma_0^2}\nonumber\\
&+&\hspace{-.74mm}\thalf B\delta \zeta^2\phi^2\!\!\left[\sigma^2
+\vmg{\pi}^2-\frac{\phi^2}{2\zeta^2}\right]-\ttquar\epsilon_1'\label{lm}\\
&-&\tquar\epsilon_1'\left(\frac{\phi}{\phi_0}\right)^{\!2}
\left[\frac{4\sigma}{\sigma_0}-2\left(\frac{\sigma^2
+\vmg{\pi}^2}{\sigma_0^2}\right)-\left(\frac{\phi}{\phi_0}\right)^{\!2}
\,\right]\,.  \nonumber
\end{eqnarray}
Here $\sigma$ and $\vmg{\pi}$ are the chiral fields, $\phi$ the
dilaton field, $\omega_{\mu}$ the vector meson field and ${\bf
b}_{\mu}$ the vector-isovector $\rho$-meson field, introduced here in order to study 
asymmetric nuclear matter. The field strength
tensors are defined in the usual way
$F_{\mu\nu}=\partial_{\mu}\omega_{\nu}-\partial_{\nu}\omega_{\mu}$,
$\vm{B}_{\mu\nu}=\partial_{\mu}\vm{b}_{\nu}-\partial_{\nu}\vm{b}_{\mu}$.
The vacuum mean field value of the scalar fields is:
$\bar {\phi} ={\phi_0}$, $\bar {\sigma}={\sigma_0}$ and
$\bar {\vmg{\pi}}=0$. Moreover,
$\zeta={\phi_0}/{\sigma_0}$, B and $\delta$ are constants and
$\epsilon_1'$ is a term which breaks explicitly the chiral invariance
of the lagrangian. 

It is easy to check that the first two terms of the potential ${\cal V}$
and the chiral symmetry breaking terms contribute to the trace anomaly, 
so that eq.~(\ref{schechter}) is satisfied with 
$\epsilon_{\rm vac}=-{\tquar}B{\phi_0}^4(1-{\delta})-\epsilon_1'$, while the third term fixes the vacuum 
conditions. Moreover, since $\epsilon_{\rm vac}$ takes contributions both
from gluons and quarks, 
the choice $\delta = \frac{4}{33}$ is suggested by the form of the QCD beta-function
at first loop, Eq.~(\ref{beta}).

Notice that, as in the case of the well known "Mexican hat" potential,
the minimum of the potential ${\cal V}$ is degenerate if
$\epsilon_1'=0$ and the ground state satisfies the "chiral circle"
equation $(\sigma^2+\pv^2)/{\sigma_0}^2=\phi^2/{\phi_0}^2$.  Notice
therefore that when the mean value of the dilaton field drops to zero
also the radius of the chiral circle drops to zero. In other words,
the restoration of scale invariance implies the vanishing of all the
condensates, including the chiral one.

It is important to remark that
scale invariant masses for the vector mesons can be generated by coupling the vector
fields either to the dilaton or to the chiral fields.
In the initial versions of
the model \cite{Ellis:1992ey,Heide:1992tk} the coupling 
$\thalf\left[G_{\omega\sigma}(\sigma^2+\pv^2) \right]\omega_\mu\omega^\mu$ 
was adopted, which leads to the scenario proposed by Brown and Rho where 
the vector meson masses scale as the mass of the nucleon \cite{Brown:1991kk}.
As it will be clarified in the following, that choice leads to an anomalous behavior
of the sigma mean field which increases at very large densities.
In Ref.~\cite{Heide:1993yz} a more general 
combination $\thalf\left[G_{\omega\phi}\phi^2+G_{\omega\sigma}(\sigma^2+\pv^2) 
\right]\omega_\mu\omega^\mu$ was tested, but that choice 
does not provide a good phenomenology for nuclei.
In the lagrangian of Eq.~(\ref{lb}) the $\omega$ and $\rho$ masses are generated
by coupling the vector mesons only to the dilaton field so that
in the vacuum $m_{\omega}=G_{\omega\phi}^{1/2} \phi_0$ and
$m_{\rho}=G_{\rho\phi}^{1/2} \phi_0$. This choice, successfully adopted in
Refs.~\cite{Carter:1995zi,Carter:1996rf,Carter:1997fn}, does not lead to
the scenario proposed by Brown and Rho \cite{Brown:1991kk}.

The values of the parameters which we used for calculations 
are listed in Table~\ref{tab}. 
They were determined in Ref.~\cite{Carter:1995zi} by fitting 
the properties of nuclear matter and finite nuclei. 
In Ref.~\cite{Carter:1997fn} two parameter sets are provided, one with $G_4=0$ and
the other with $G_4\neq 0$.
Since a better phenomenology was obtained with the non-zero 
value of $G_4$, we use that parameter set in our calculations. 
The value of the coupling constant of the $\rho$-meson to the nucleon $g_{\rho}$
has been chosen in order to obtain for the symmetry energy at $\rho_0$
the value $E_{sym}=35$ MeV. The $\rho$-meson did not appear in Ref.~\cite{Carter:1997fn} 
where only isospin symmetric 
nuclear matter has been studied.
\begin{table}[t]
\begin{center}
\caption{Values of the parameters.}\label{tab}
\begin{tabular}{|l|c|c|} \hline
Quantity& Value\\ \hline
$G_4/g_{\omega}$& 0.19\\
$|\epsilon_{\rm vac}|^{1/4}$ (MeV)& 228\\
$g$& 9.2\\
$g_{\omega}$& 12.2 \\
$g_{\rho}$& 8.1\\
$\zeta=\phi_0/\sigma_0$& 1.41\\
$\sigma_0$ (MeV) & 102\\
${\epsilon_1'}^{1/4}$ (MeV)& 119 \\ \hline
\end{tabular}
\end{center}
\end{table}

\subsection{Thermal fluctuations and their resummation}
\label{thermal_flu}
The main purpose of our work is to study the EOS of nuclear matter at finite
temperature. When the temperature is large enough to become comparable 
to the masses of the meson fields it is mandatory to take into account the
thermal fluctuations of those fields. In the present work we 
consider the thermal fluctuations of the chiral fields and of the 
vector mesons. The fluctuations of the dilaton
field are not taken into account in the present paper, due to the large mass of the
glueball (about 1.6 GeV). In a future analysis we will consider also the thermal
fluctuations of the dilaton field, which become important when studying in detail
the temperature-density region where scale invariance is restored.

To take into account thermal fluctuations one has to resort to a technique 
(developed in Refs.\cite{Carter:1996rf,Carter:1997fn})
which allows to deal with non-polynomial functions of the fields.
The technique is the following:

a) the chiral invariant combination of the chiral fields is expanded
in mean field value plus fluctuations, as:
\begin{equation}
\frac{\sigma^2+\pv^2}{\sigma_0^2}=\frac{1}{\sigma_0^2}(\bar{\sigma}^2
+2\bar{\sigma}\Delta\sigma+\Delta\sigma^2+\pv^2)\, ,
\end{equation}
where $\bar{\sigma}$ is the sigma mean field value and $\Delta\sigma$
is its fluctuation.
The pion field has a vanishing mean field value
and therefore $\pv^2$ is also the fluctuation squared;

b) in order to simplify the evaluation of the thermal averages of the 
fluctuations, a 
crucial assumption is made \cite{Carter:1996rf,Carter:1997fn}: 
\begin{equation}
\langle\Delta\sigma^2\rangle\approx\langle\pi_i^2\rangle \, .\label{approx}
\end{equation}
This approximation is based on the observation that at low temperature the
fluctuations are small (and therefore the error in assuming the validity of
eq.~(\ref{approx}) is also small), while at large temperature chiral symmetry is
(at least partially) restored and eq.~(\ref{approx}) is a sensible approximation.
We end up therefore with a unique quantity describing all the thermal fluctuations
of the chiral fields,~$\langle\psv^2\rangle$, where $\psv^2=(\Delta\sigma^2+\pv^2)/\sigma_0^2$;\\

c) a generic function of the chiral invariant expression is expanded as a 
series of the field fluctuations and the fluctuations in $\psv^2$ are
taken to all orders~\footnote{the expansion in $\nu\Delta\nu$ is truncated 
at fourth order, because at low temperature $\Delta\nu$
is small since the mass of the sigma field is large, while at large temperature
the sigma field is small because chiral symmetry is (at least approximately)
restored.};

d) to evaluate the numerical value of $\langle\psv^2\rangle$ one needs the
thermal averages of the fluctuations squared of the chiral fields, which
are computed as integrals of the 
statistical functions:
\bqr
\langle\pi_a^2\rangle&=&\frac{1}{2\pi^2}\int\limits_0^\infty\,dk
\frac{k^2}{e_{\pi}}\frac{1}{e^{\beta (e_{\pi}-\mu_{a})}-1}\label{thep}\\
\langle\Delta\sigma^2\rangle&=&
\frac{1}{2\pi^2}\int\limits_0^\infty\,dk
\frac{k^2}{e_{\sigma}}\frac{1}{e^{\beta e_{\sigma}}-1}\;. \label{the}
\eqr

e) finally, the series in $\langle\psv^2\rangle$ is resummed.

In a more recent paper \cite{Mocsy:2004ab} a new technique to
evaluate thermal fluctuations has been developed, in which the fluctuations
of $\sigma$ and $\pv$ are always considered as independent quantities. 
In particular, a generic function of $\sigma$ and $\pv$ is first expanded
as a series of the chiral fields fluctuations and these fluctuations are 
taken to all orders. The series in $\langle\Delta\sigma^2\rangle$ 
and $\langle\pv^2\rangle$ is then resummed. In the case of isospin
symmetric matter the fluctuations of the three components of the pionic field are equal and 
the following result is obtained:
\begin{equation}
\langle f(\bar{\sigma}+\Delta\sigma,\pv^2)\rangle=\int_{-\infty}^{\infty}dz P_{\sigma}(z)
\int_{0}^{\infty}dy y^2 P_{\pi}(y)f(\bar{\sigma}+z,y^2)\label{mish}
\end{equation}
where
\begin{eqnarray}
P_{\sigma}(z)=(2\pi\langle\Delta\sigma^2\rangle)^{-1/2}
\exp\left(-\frac{z^2}{2\langle\Delta\sigma^2\rangle}\right)\nonumber\\
P_{\pi}(y)=\sqrt{\frac{2}{\pi}}\left(\frac{3}{\langle\pv^2\rangle}\right)^{3/2}
\exp\left(-\frac{3y^2}{2\langle\pv^2\rangle}\right)
\end{eqnarray} 
are the gaussian weighting functions.
This procedure can be generalized to the case of asymmetric matter, 
where the fluctuations of the pionic field are independent. In that case
instead of the bidimensional integral of Eq.~(\ref{mish}) a four dimensional
integral is obtained.
Notice that the procedure developed 
in Ref.~\cite{Mocsy:2004ab} reduces to the technique discussed in 
Refs.~\cite{Carter:1996rf,Carter:1997fn} if 
the thermal fluctuations of the chiral fields are assumed to be all equal.

In the present work we have mainly adopted the technique of Refs.\cite{Carter:1996rf,Carter:1997fn}
because our main aim is to extend the calculations towards asymmetric nuclear matter and
astrophysics and we like to be able to directly compare our results with the ones
obtained in previous calculations. Most of the results discussed in the following
are rather independent from the technique adopted: e.g. the value of the sigma 
mean field is the same in both cases. Nevertheless it is necessary to use the
technique developed in Ref.\cite{Mocsy:2004ab} when computing the
isospin splitting of quantities strongly dependent on the thermal 
fluctuations, in particular the isospin splitting of the mass of the pion.\\

\subsection{Thermodynamics}

The value of the fields and of the various thermodynamical
quantities at finite temperature and density can
be obtained in two ways: i) by solving the field equations (obtained by 
minimizing the action) and substituting the solutions into the thermodynamical
functions
or, ii) by minimizing directly the relevant thermodynamical quantities,
such as the free energy (if working at given density and temperature), or
the thermodynamical potential (if working at given chemical potential and temperature)
or the energy density (if working at given entropy and density) \cite{Kapusta:2006pm}.
These two approaches are in principle equivalent, although the approximations discussed
in the previous subsection can introduce some (small) discrepancy \cite{Carter:1997fn}.
In the present paper
we have decided to minimize directly the thermodynamical quantities because in future works
we aim at introducing quark and gluon degrees of freedom. Since in Ref.\cite{Drago:2001gd}
the pure gluonic sector was described in terms of an effective thermodynamical
potential (without introducing a model lagrangian), the second approach is the 
appropriate one if quarks and gluons are introduced following the scheme proposed in 
\cite{Drago:2001gd}.

The thermodynamical potential, within the mean field approximation and
taking into account the thermal fluctuations, reads:
\begin{widetext}
\bqr
\frac{\Omega}{V}\pj\,&=&\,\pj\langle{\cal V}\rangle
-\thalf m_\omega^{2}\chi^2\omega_0^2-\thalf m_\rho^{2}\chi^2 b_0^2 
- G_4^4\omega_0^4
-\thalf m_{\sigma}^{*2}\langle\Delta\sigma^2\rangle
-\thalf m_{\pi}^{*2}\langle\vmg{\pi}^2\rangle\nonumber\\
&+&\frac{T}{2\pi^2}\int dk\,k^2
\left[\ln(1-e^{-\beta e_{\sigma}})
+3\ln(1-e^{-\beta e_\omega})\right]\nonumber\\
&+&\frac{T}{2\pi^2}\int dk\,k^2
\left[\sum_{a=1,3}\ln(1-e^{-\beta (e_{\pi}^*-\mu_{\pi_a}^*)})
+3\sum_{a=1,3}\ln(1-e^{-\beta (e_\rho^*-\mu_{\rho_a}^*)})\right]\nonumber\\
&-&\frac{T}{\pi^2}\sum_{i=p,n}\int{\rm d}k_i\,k_i^2\left[
\ln\left(1+e^{-\beta(E_i^*-\mu_i^*)}\right)
+ \ln\left(1+e^{-\beta(E_i^*+\mu_i^*)}\right)\right] \, ,
\label{grand}
\eqr
\end{widetext}
where the quantities into brackets are thermal averages.
We omit to show explicitly 
the complicate expression for $\langle V \rangle$, which can be 
obtained using the procedure described in the previous subsection
and is given in Eq.~(21) of Ref.\cite{Carter:1996rf}.
The subscripts n and p stay for proton and neutron, respectively,
the index $a$ runs on the three isospin components of the pion and of the $\rho$-field and,
finally, $\chi=\phi/\phi_0$.

The energies which appear into the thermodynamical integrals read 
$e_i^*=\sqrt{{m^*_i}^2+k^2}$ where $m^*_i$ is the
effective mass of the i{\it th}-field, which can be obtained by computing
the thermal average of the
second derivative of the lagrangian with respect to the field:

\begin{widetext}
\begin{eqnarray}
M_{N}^*&=&g\langle\sqrt{{\sigma}^2+{\vmg{\pi}}^2}\rangle
\,,\\[2mm]
m_\omega^{*2}&=&  m_\omega^2\chi^2 \label{massaomega}
\,,\\[2mm]
m_\rho^{*2}&=&  m_\rho^2\chi^2 \label{massarho}
\,,\\[2mm]
\sigma_0^2m_{\sigma}^{*2}&=&(B_0\delta+\epsilon_1')\chi^2  
+\left\langle-\frac{B_0\delta\chi^4\sigma_0^2}{\sigma^2+\vmg{\pi}^2}
+\frac{2B_0\delta\chi^4\sigma_0^2\sigma^2}{(\sigma^2+\vmg{\pi}^2)^2}
\right\rangle + 
g\so^2(\rho_{S p}+\rho_{S n})\left\langle\frac{1}{\sqrt{\s^2+\pv^2}} -
\frac{\s^2}{(\s^2+\pv^2)^{3/2}} \right\rangle\,,
\label{massasigma}\\[2mm]
\sigma_0^2m_{\pi}^{*2}&=&(B_0\delta+\epsilon_1')\chi^2
+\left\langle-\frac{B_0\delta\chi^4\sigma_0^2}{\sigma^2+\vmg{\pi}^2}
+\frac{2B_0\delta\chi^4\sigma_0^2\pi_a^2}{(\sigma^2+\vmg{\pi}^2)^2}
\right\rangle
+ g\so^2(\rho_{S p}+\rho_{S n})\left\langle\frac{1}
{\sqrt{\s^2+\pv^2}}-
\frac{\pi_a^2}{(\s^2+\pv^2)^{3/2}} \right\rangle\,,
 \;\label{massapione}
 \end{eqnarray}
\end{widetext}
where $\rho_{S p}$ and $\rho_{S n}$ are the scalar densities of the protons and 
of the neutrons, respectively.
As already remarked above, we have omitted the thermal average on the dilaton field because its
thermal fluctuations are suppressed due to the large mass of this field.
Notice that the masses of the chiral fields depend 
on their thermal fluctuations, which in turn depend on the masses.
Due to these complicate dependencies, the minimization of the relevant thermodynamical
quantities has to be done iteratively.

Finally, the chemical potentials entering the thermodynamical functions are the
effective ones. In the case of the
nucleons, the effective chemical
potentials are related to the standard ones as follows:
\begin{eqnarray}
\mu_p^*\,&=&\, \mu_p-g_{\omega}\omega_0-\thalf
g_{\rho}b_0\nonumber\\
\mu_n^*&\,=&\, \mu_n-g_{\omega}\omega_0+\thalf
g_{\rho}b_0\;. \label{potchim}
\end{eqnarray} 
Notice that, when asymmetric nuclear matter is considered, the charged
isovector mesons acquire a non zero chemical potential, due to the conservation 
of the isospin charge. The chemical potentials are determined by the equations: 
\begin{eqnarray}
\mu_{\pi^+} = \mu_{\rho^+}  = \mu_{p}-\mu_{n}  
\end{eqnarray}
and the effective ones, which enter e.g. Eq.~(\ref{grand}), are
given by the relation:
\bq
\mu_{\pi^+}^*= \mu_{\rho^+}^*=\mu_{p}^*-\mu_{n}^*= \mu_{p}-\mu_{n}-g_{\rho}b_0\, .
\eq

\subsection{Goldstone theorem at finite temperature}
\label{goldstone}
When discussing a field theory at finite temperature, one has to choose an
approximation scheme which do not violate the symmetries of the lagrangian.
In particular, adopting a chiral lagrangian
and working in the chiral limit, at T=0 the pion is massless
due to the Goldstone theorem. At finite temperature though, it is not automatic
that the Goldstone theorem is satisfied, since a wrong truncation can spoil
the theorem.
There are several contributions to the various observables
which one can in principle take into account:\\
a) mean field contributions, neglecting all fluctuations;\\
b) fluctuations in $\pv^2$ and $\sigma^2$, particularly at finite temperature;\\
c) fermionic contributions associated with the so-called particle-hole 
excitations;\\
d) further corrections, as the exchange diagrams discussed e.g. 
in Ref.\cite{Mocsy:2004ab} at the end of Sec.~IVA.

A important guideline in deciding which truncation scheme is
appropriate (in other terms, a way for deciding that including a further
correction is not necessarily a better approximation) is to check what
happens to the Goldstone theorem at finite temperature.   
For instance, in \cite{Carter:1996rf} (which describes the same
approximation scheme adopted here) the Goldstone
theorem is satisfied also at finite temperature by
taking into account contributions a) and b) and neglecting the others 
(see Fig.~3 of Ref.~\cite{Carter:1996rf} and Fig.~\ref{mpichi0} of the present paper). It is well
known that, when studying chiral models, another truncation exists which
allows the Goldstone theorem to be satisfied at finite temperature and
it corresponds to
the inclusion of contributions a) and c) (see e.g. \cite{Kapusta:2006pm}). In the chiral
limit one can check that indeed the contributions a) and c) to the mass mutually cancel.  
On the other
hand, if one includes a) and b) and c), the Goldstone theorem is {\it not}
satisfied at finite temperature, as already discussed in Ref.~\cite{Mocsy:2004ab}.

In conclusion, one has to resort either to
a) + b) (as we are doing here) or to a) + c) (as discussed e.g. in \cite{Kapusta:2006pm}).  
If one wants to add c) to
a) and b), other contributions, as the ones in d) has also to
be considered in order to satisfy Goldstone theorem.

We will come back to this problem in Sec.\ref{bose}, when discussing the possibility that the pions
form a Bose-Einstein condensate.

\section{Matter at high baryon density and temperature}
\label{lab}

In this section we will focus on 
nuclear matter calculations at rather high temperature and relatively low
isospin asymmetry. These calculations are aimed at providing previsions on the 
EOS of matter in regions of the energy-density vs. baryon-density 
plane which can be tested with HIC experiments. 
Those experiments are particularly interesting at energies of the order 
of a few ten A GeV, which correspond to the regime at which maximum baryon
densities can be reached. In this Section we first discuss the restoration
of chiral symmetry and of scale invariance in various region of the 
energy-density vs. baryon density plane, we then analyze various 
phenomenological implications
of the partial restoration of these two symmetries.

\subsection{Phase diagrams}
The advantage of using a sophisticated model which includes both 
the chiral dynamics and the gluon 
condensate dynamics, is that one can 
make previsions on the restoration of both the chiral and the scale symmetry.
These previsions are different if one considers the case of exact chiral symmetry 
or the case where a symmetry breaking term is present into the
lagrangian. In the following we study both cases.

\subsubsection{Chiral restoration in the chiral limit ($\epsilon_1'=0$)}

\begin{figure}[t!]
\centering
\includegraphics[scale=0.35]{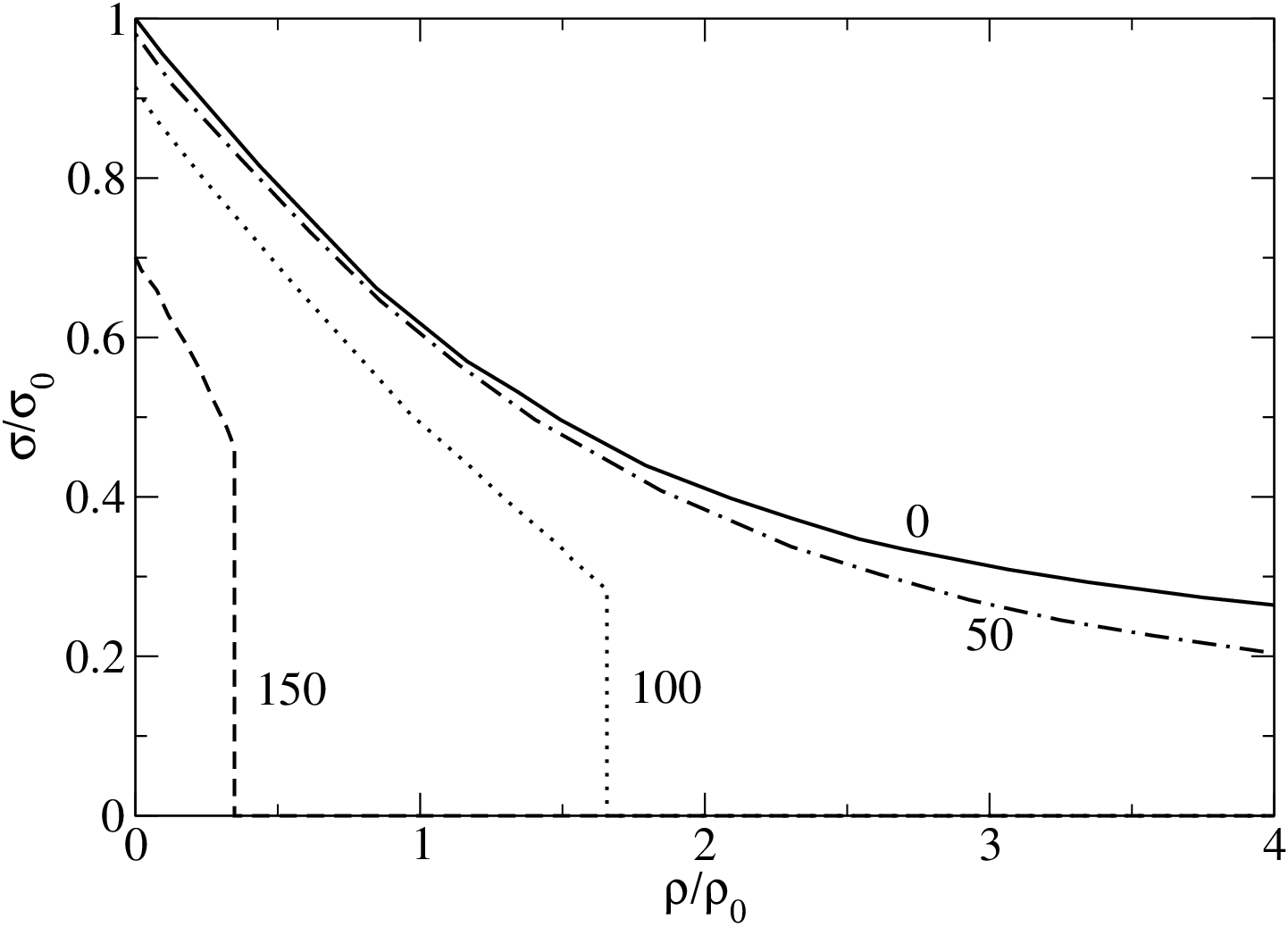}
\caption{
\label{sigmachi0} \footnotesize
Mean value of the sigma field as a function of the baryon density
for different values of the temperature. Here $\epsilon_1'=0$.}
\end{figure}

\begin{figure}[t]
\centering
\includegraphics*[scale=0.35]{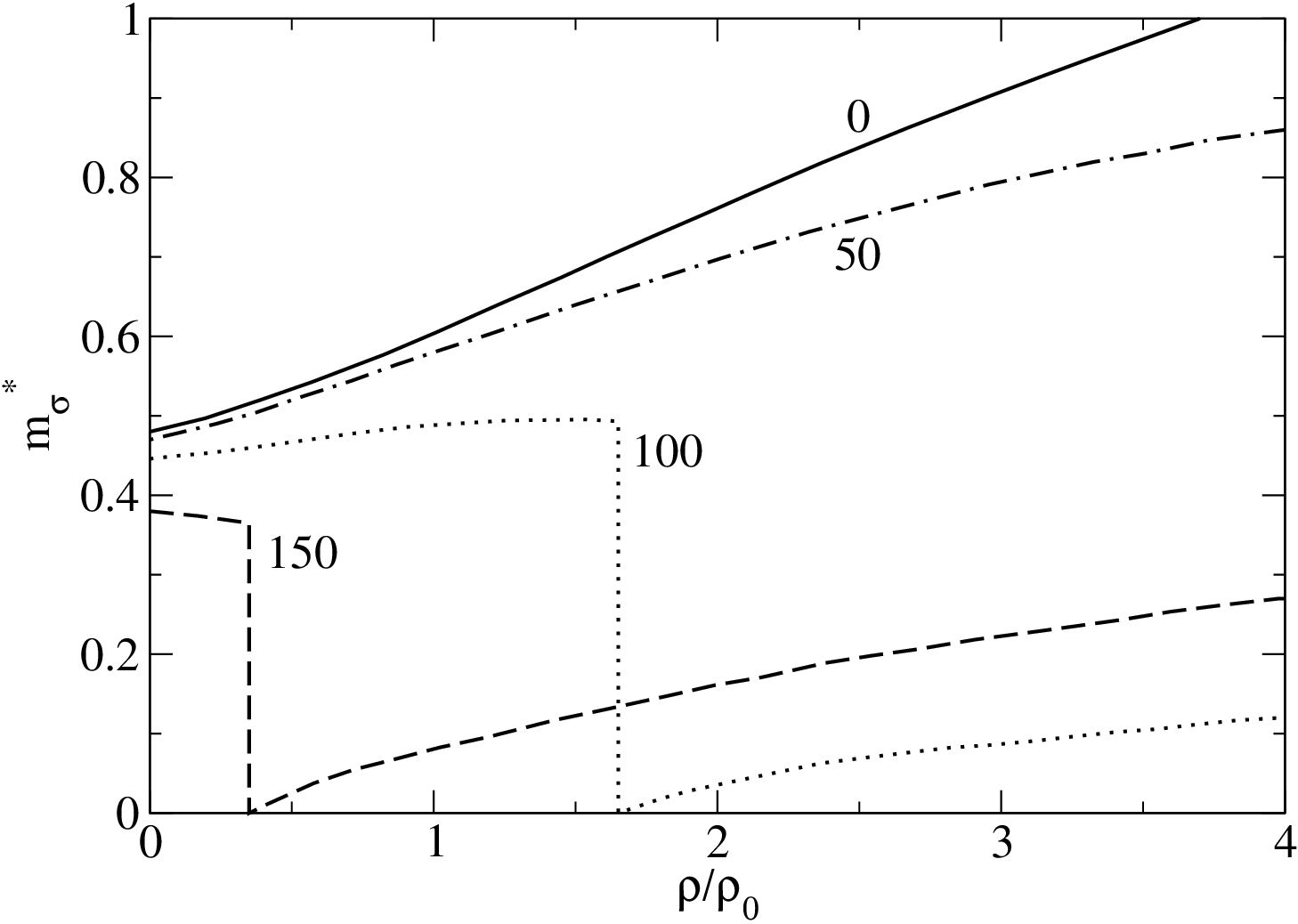}
\caption{
\label{msigchi0} \footnotesize 
Effective sigma meson mass as a function of the baryon density
for different values of the temperature. Here $\epsilon_1'=0$.}
\end{figure}

\begin{figure}[tb]
\centering
\includegraphics*[scale=0.35]{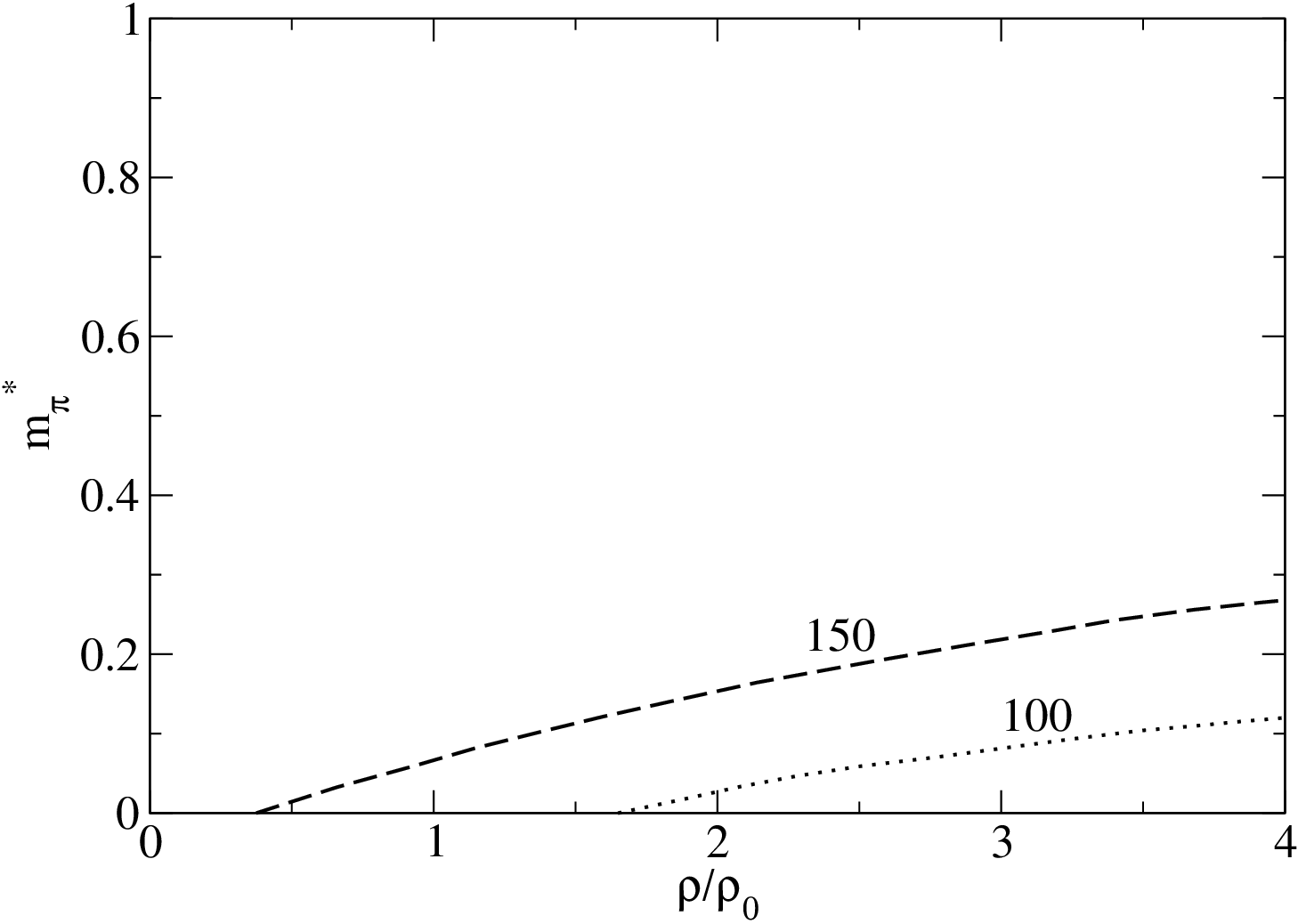}
\caption{
\label{mpichi0} \footnotesize 
Effective pion mass as a function of the baryon density
for different values of the temperature. Here $\epsilon_1'=0$.}
\end{figure}

\begin{figure}[tb]
\centering
\includegraphics*[scale=0.35]{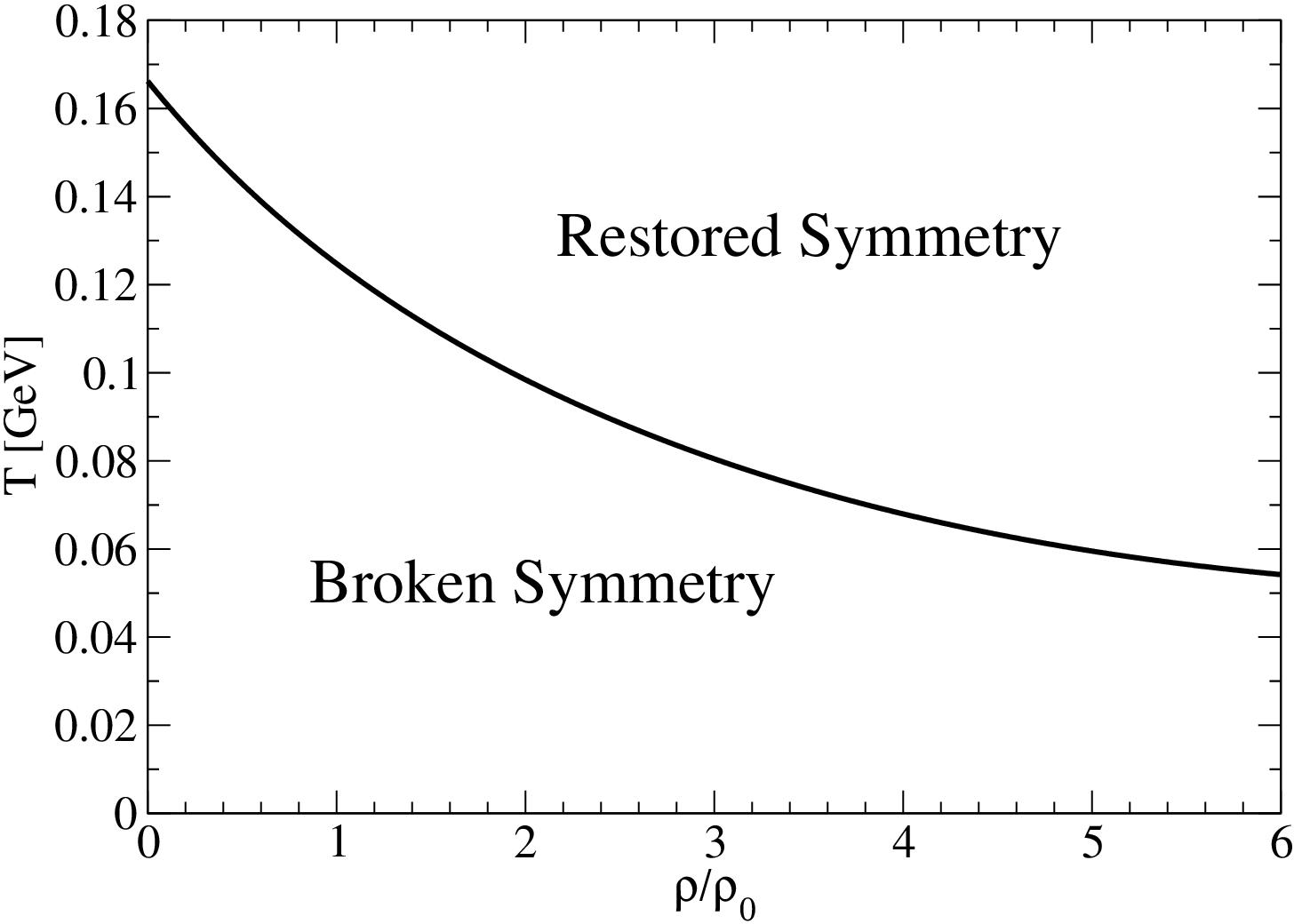}
\caption{
\label{trans} \footnotesize 
Chiral phase diagram for symmetric nuclear matter in the temperature 
versus baryon density plane. Here $\epsilon_1'=0$.}
\end{figure}

The restoration of chiral symmetry at finite density and temperature.
occurs as a first order transition only when the chiral invariance of the 
lagrangian is exact, or 
equivalently, only when the explicit symmetry breaking term $\epsilon_1'=0$.
In this case, at a given temperature we observe that for densities larger 
then a critical one 
the mean field value of 
$\sigma$ drops to zero and the 
masses of the chiral fields become degenerate.
In Fig.~\ref{sigmachi0} the behavior of the $\sigma$ field, as a function of the density and for 
various temperatures is displayed. The transition turns out to be first order, due to the 
discontinuous behavior of the sigma field. Similarly, in Figs.~\ref{msigchi0},~\ref{mpichi0}
we show the effective masses of the chiral fields, which become degenerate when the symmetry is
restored.    
The chiral phase diagram for isospin symmetric nuclear matter is shown in Fig.~\ref{trans} in the 
temperature versus baryon density plane. 
At zero baryon density the temperature at which the chiral
restoration occurs is $T_C=165$ MeV, which is 
rather close to the QCD critical temperature estimated by lattice 
QCD \cite{Cheng:2006aj,Cheng:2006qk}. Obviously,
in the case of QCD the restoration of chiral symmetry comes together with the 
appearance of the quark-gluon plasma phase.
As it can be observed, the critical temperature decreases with 
increasing density, 
but it never reaches zero, not even at very large densities. 
Numerically it tends asymptotically to $\sim$ 40 MeV.
It can be interesting to notice that some (very preliminary)
lattice calculations at finite density seem to support this result~\cite{Blum:1995cb}.
The chiral phase restoration in the chiral limit is also represented in Fig.~\ref{trans2}, 
in the energy-density vs. baryon-density plane, by two closely spaced solid lines.
Since the chiral transition is first order
these lines are separated by a small
energy gap. 
Indeed, at finite temperature a first order transition occurs with a
jump in the energy density and in the entropy.\\ The temperatures
investigated in this Subsection are not large enough for the dilaton
field to change significantly from its zero temperature value,
therefore we could decouple the discussion of chiral symmetry
restoration from the restoration of scale invariance. This simplified
scenario will not hold in the more realistic $\epsilon_1'\neq0$ case,
discussed in the next Subsection.

\subsubsection{Chiral restoration with a chirally broken lagrangian,~$\epsilon_1'\neq0$}

\begin{figure}[b!]
\centering
\includegraphics*[scale=0.35]{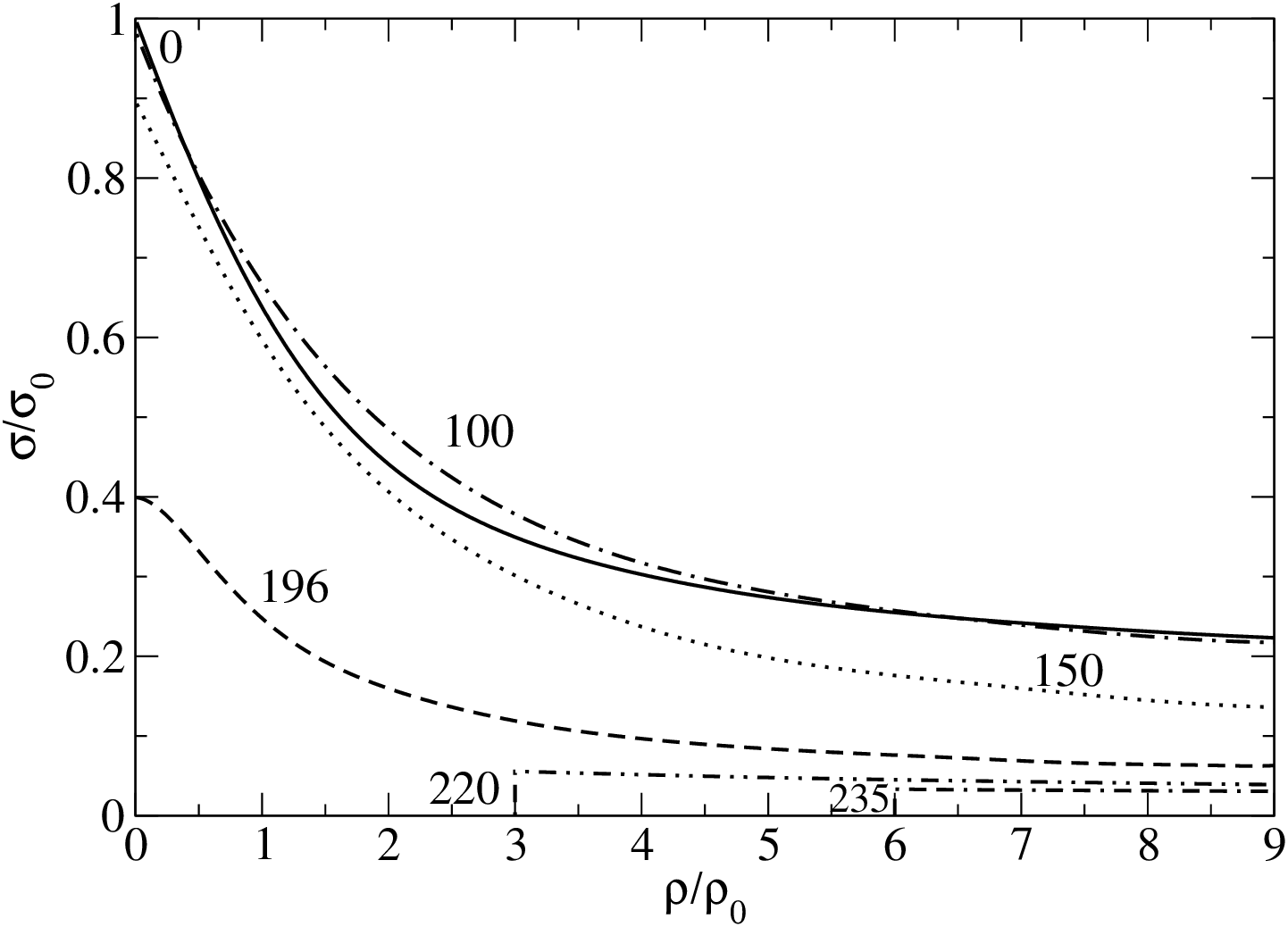}
\caption{
\label{sigmachi} \footnotesize
Same as in Fig.(\ref{msigchi0}), but for the case where
the chiral symmetry is explicitly broken ($\epsilon_1'\neq 0$).}
\end{figure}

\begin{figure}[t!]
\centering
\includegraphics*[scale=0.35]{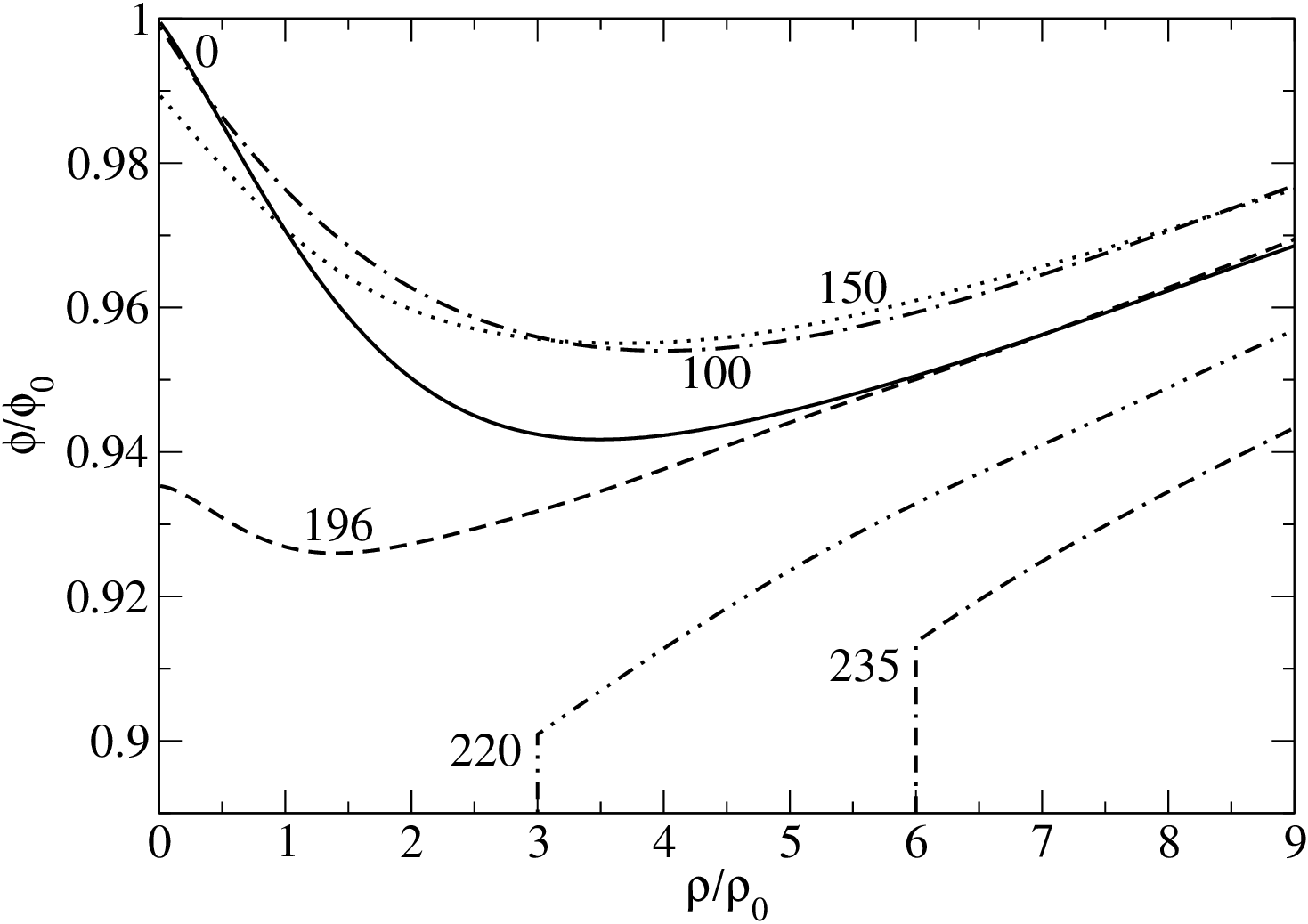}
\caption{
\label{chi} \footnotesize
Mean value of the dilaton field as a function of the baryon density
for different values of the temperature. Here $\epsilon_1'\neq 0$.}
\end{figure}

\noindent In the more realistic case in which $\epsilon_1'\neq 0$ and
the pion has a finite mass, we observe that the $\sigma$ mean field is
continuous everywhere and it never drops to zero, at any density as
long as T~$\lesssim 196$ MeV, see Fig.~\ref{sigmachi}.  Therefore a
chiral phase transition, as that observed in the chiral limit case,
does not occur.  On the other hand, the $\sigma$ mean field decreases
continuously with the density and drops rapidly when the temperature
raises from T~$\sim 100$ MeV to T~$\sim 196$~MeV reaching very small
but not vanishing values.  This behavior, called ``cross over'', is
predicted by QCD lattice calculations with two flavors at zero or
small densities.\\ At T~$\gtrsim 196$ MeV the dilaton field drops to
zero (initially in the small density region, see Fig.~\ref{chi}) and
scale invariance (and therefore also chiral symmetry) is restored, as
discussed in the next section.\\ The large density behavior of the
$\sigma$ mean field can be obtained by solving its equation of motion,
what can be done analytically for $T=0$. For large values of the
scalar density $\rho_S$, we obtain $\sigma\propto 1/\rho_S$.  Since
$\rho_S$ scales as $\rho^{1/3}$ for large $\rho$, the $\sigma$ mean
field decreases monotonously but slowly with $\rho$ and it never
vanishes.  This analytic result remains numerically true also at
finite temperatures (for T~$\lesssim 196$ MeV).

\begin{figure}[t!]
\centering
\includegraphics*[scale=0.35]{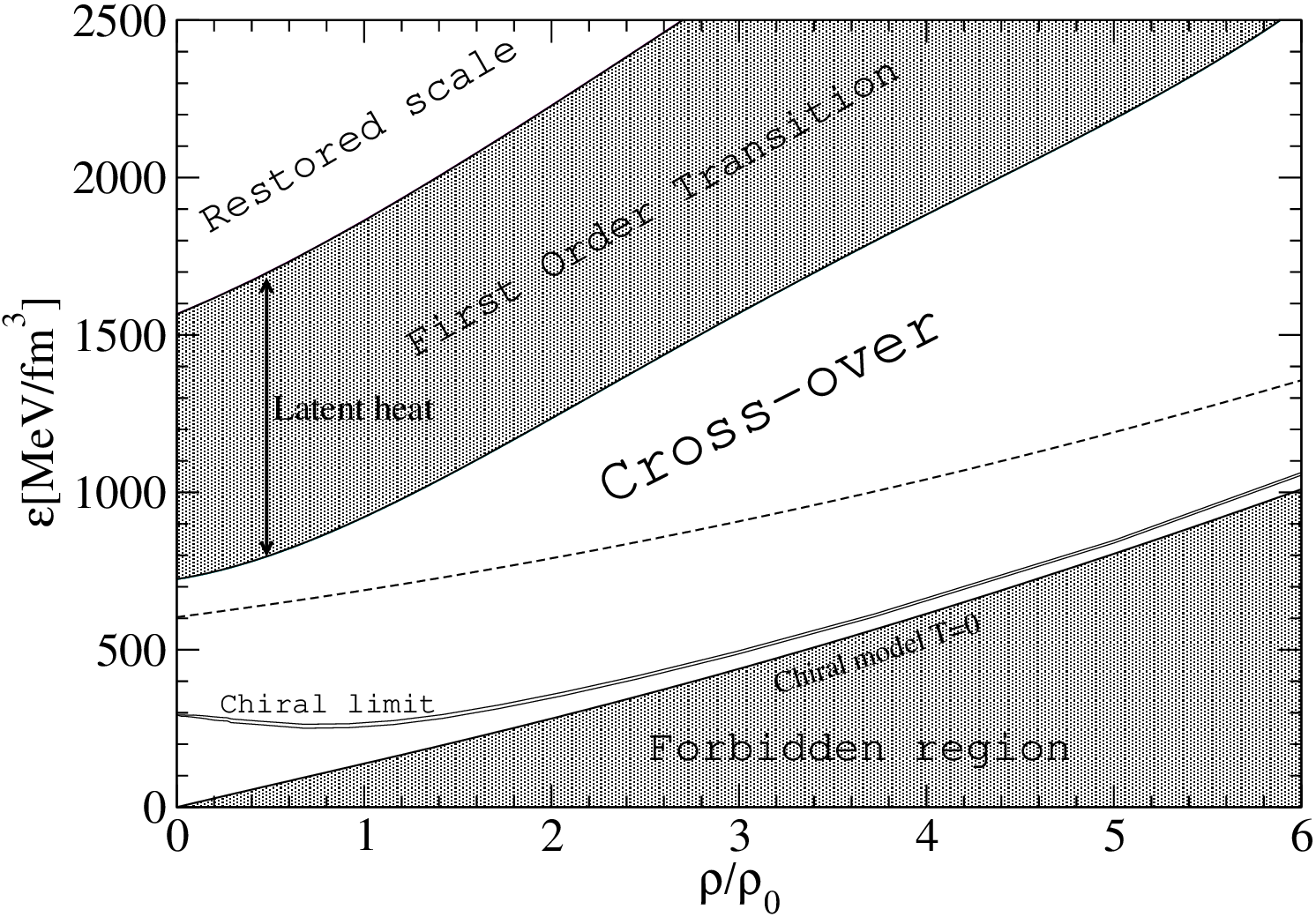}
\caption{
\label{trans2} \footnotesize 
Phase diagram for symmetric nuclear matter in the energy-density versus 
baryon-density plane.
Two closely spaced solid lines separate 
the (low-lying) region where chiral symmetry is broken from the (upper-lying) 
region where chiral symmetry is restored
in the chiral limit $\epsilon_1'=0$. All other graphical signs refer to the broken symmetry case
$\epsilon_1'\neq 0$. The lower shaded area indicates 
the forbidden region under the T=0 EOS. The upper shaded 
area separates the (low-lying) region where scale symmetry is broken from the region 
where scale symmetry is restored. The dashed line indicates the maximum (negative) variation
of the $\sigma$-field mean value with the temperature (see text).}
\end{figure}

\begin{figure}[b!]
\centering
\includegraphics*[scale=0.35]{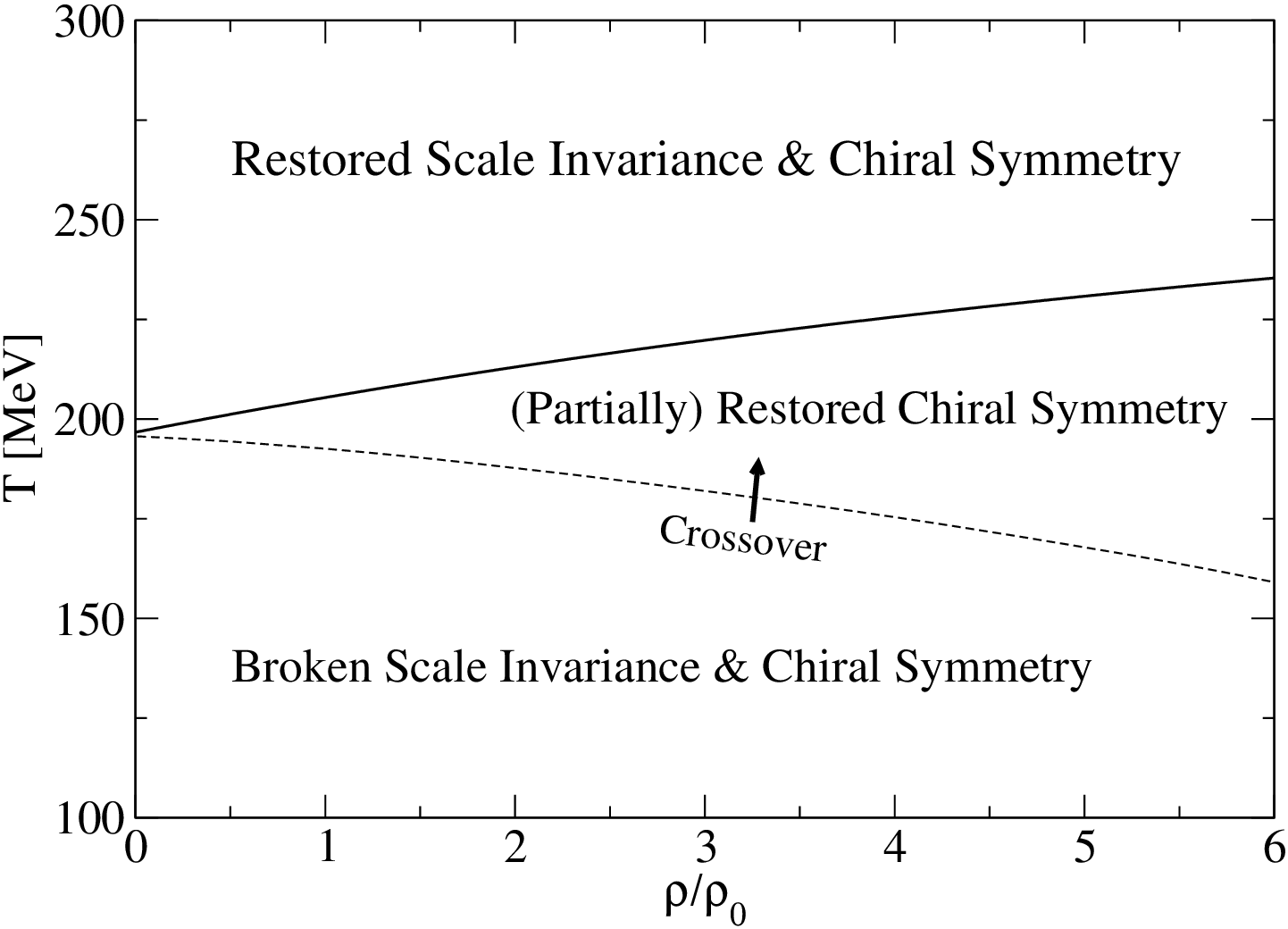}
\caption{
\label{trans3} \footnotesize 
Phase diagram for symmetric nuclear matter in the temperature versus baryon density plane.}
\end{figure}

Let us now concentrate on Fig.~\ref{trans2}, where a sort of ``phase
map'' for symmetric nuclear matter is shown in the energy-density
versus baryon-density ($\rho,\epsilon$) plane.  The region in which
the variation of the $\sigma$ field is more rapid is indicated by a
dashed curve, which is obtained by joining the points of the
($\rho,\epsilon$) plane where the modulus of the derivative of
$\sigma$, with respect to the temperature, takes its maximum value.
In Ref.~\cite{Arsene:2006vf} the maximum energy and baryon densities
reached at the moment of maximum compression during a HIC experiment
are computed within several models.  It is interesting to notice that
in all model estimates when the beam energy exceeds $\sim 5$ A GeV the
maximum energy density lies above the dashed line of Fig.~\ref{trans2}
and therefore signatures of the partial restoration of chiral symmetry
should be observable.  In particular, the softening of the EOS due to
partial chiral symmetry restoration will be discussed later in this
Section.

\subsubsection{Scale symmetry restoration}
\label{scaleres}
\noindent In the model here discussed, the restoration of scale
invariance takes place when the mean value of the dilaton field drops
to zero. As shown in Fig.~.\ref{chi} the restoration of scale
invariance takes place (in the mean field approximation) as a first
order transition, signaled by the discontinuous behavior of the dilaton
field.  A important result, shown in Fig.~\ref{trans2}, is that a
large energy gap appears between the two phases (due to the large
amount of energy released by the condensates which drop to zero). This
gap is of order of 1 GeV/fm$^{3}$, a value not too different from the
jump in energy density seen in lattice QCD in the pure gauge
sector. Clearly, a direct comparison is meaningless since in full QCD
the deconfinement transition (at zero baryon density) is a crossover
and, moreover, we do not have quarks in our model.

At zero baryon density, we obtain a critical 
temperature $T_c(\rho = 0)\sim$~196 MeV (see Fig.~\ref{trans3}), which corresponds to a jump
in energy density from 
$\epsilon_c^-(\rho = 0)=725$ MeV/fm$^3$
to $\epsilon_c^+(\rho = 0)=1565$ MeV/fm$^3$. This critical temperature is significantly lower
than the one estimated in \cite{Carter:1996rf,Carter:1997fn}. This reduction is due to the 
contribution of the vector mesons, which provide a huge contribution to the 
pressure when the scale invariance is restored (since their masses drop to zero).
For instance, in Ref.~\cite{Carter:1996rf}, 
where no vector meson is present, the scale restoration occurs 
at $T\sim$~550 MeV. In Ref.~\cite{Carter:1997fn} the presence of the
$\omega$ meson lowers 
the critical temperature to $T\sim$~300 MeV, while the inclusion 
in our calculations of the $\rho$ meson (together with $\omega$) 
lowers the restoration temperature 
to $T\sim$ 200 MeV.

We notice from Fig.~\ref{trans2} that the critical energy density
$\epsilon_c(\rho)$ increases with $\rho$. Also the critical
temperature for scale invariance restoration increases with $\rho$, as
shown in Fig.~\ref{trans3}.  This interesting effect arises from the
repulsive contribution of the $\omega$ vector meson, which contributes
to the energy density as $E_\omega=(1/2)(g_\omega/m_\omega^*)^2
\rho^2$ (here for simplicity we have assumed $G_4=0$).  Since the
$\omega$ meson mass scales with the dilaton field (see
Eq.~(\ref{massaomega})), $E_\omega$ increases when the dilaton field
decreases. Therefore at large $\rho$ it is less convenient to restore
scale symmetry~\footnote{Following similar arguments, if $m_{\omega}$
  and $m_{\rho}$ are generated by coupling the vector mesons to the
  chiral fields, as discussed in subsection~\ref{lagrangian}, the
  $\sigma$ mean field increases at large $\rho$. Such a behavior was
  indeed obtained in Refs.~\cite{Ellis:1992ey,Heide:1992tk}.}. This
effect is also clearly visible in Fig.~\ref{chi} which shows how the
dilaton field increases again at large densities.

\noindent
A few caveats are in order:
\begin{itemize}
\item
we are working in a mean field approximation
and it is well known that the order of a phase transition cannot be 
determined in that approximation;
\item
at the moment we have not included
the thermal fluctuations of the dilaton field, while we have taken into account
the thermal fluctuations of the vector mesons. The reason for this choice is that
the contribution of the vector mesons largely dominates the contribution
of the scalar dilaton field, since the contribution of the vector (isovector) fields is
multiplied by their degeneracy. Moreover, to take into account the dilaton 
thermal contribution one has to overcome the difficulties associated with the logarithmic
term. An expansion and resummation similar to the one performed for the chiral fields should
be done, what we plan to do in a future work.
\end{itemize}

\subsection{Bulk modulus}

\begin{figure}[ct!]
\centering
\includegraphics*[scale=0.35]{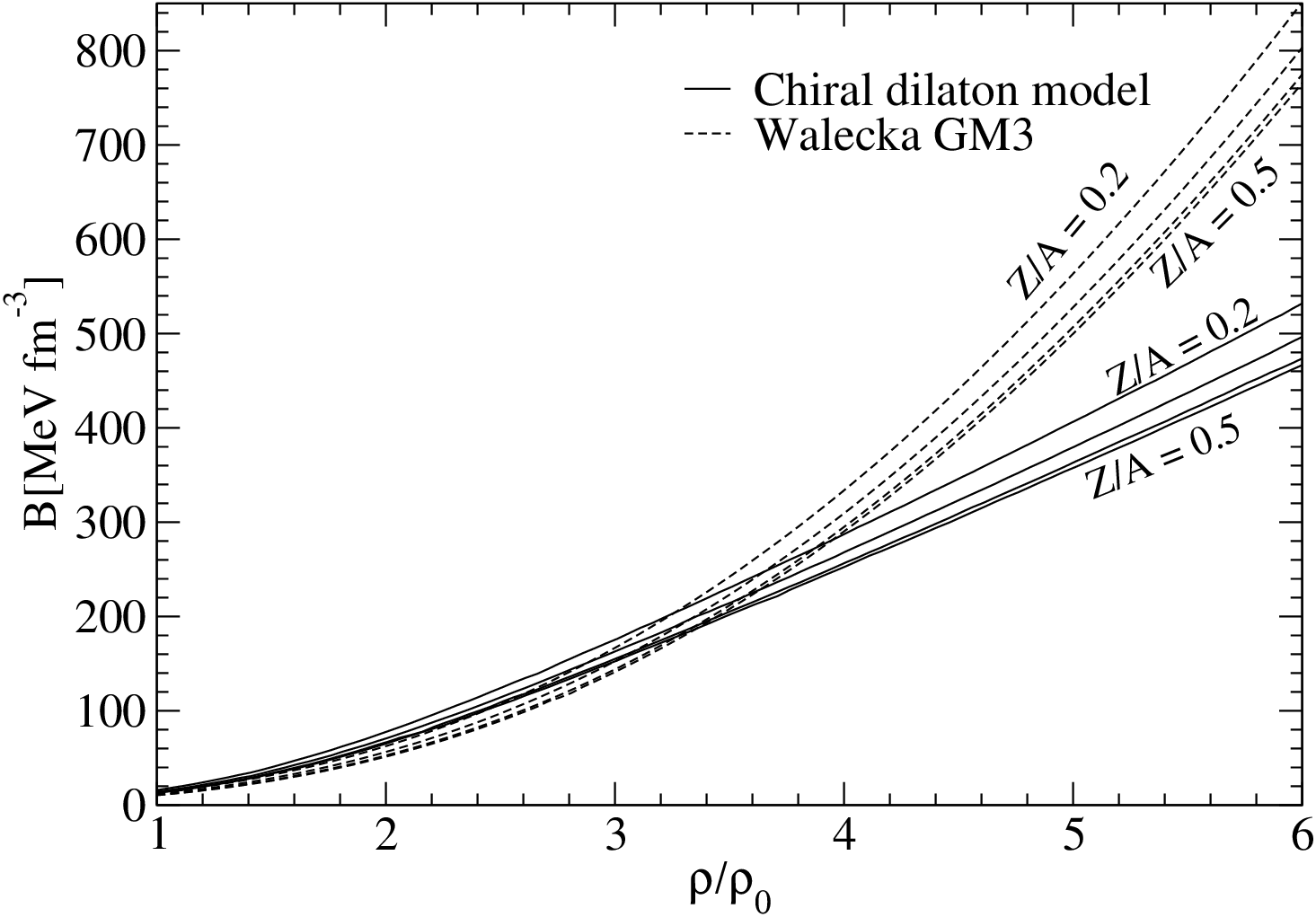}
\caption{\label{bulk} 
\footnotesize Bulk modulus $B=\rho \partial{P}/\partial{\rho}$ 
as a function of the baryon density at T=50 MeV for both the CDM and the Walecka GM3 models. 
Here various values of the proton ratio are considered.}
\end{figure}

Here we discuss bulk properties of matter within the CDM. In particular 
we show in Fig.~\ref{bulk} the value of 
the bulk modulus $B=\rho \,\partial{P}/\partial{\rho}$ as a function of 
$\rho$ at T=50 MeV. 
Here the calculations are performed for various isospin densities
and we compare with the same quantity computed
within the Walecka GM3 model \cite{Glendenning:1991es}. 
Notice that the bulk modulus computed with 
the CDM is larger than the one calculated with GM3 
when $\rho \lesssim 3 \,\rho_0$. This is due to the larger 
value of the CDM incompressibility parameter at $\rho_0$
($K^{-1}=9\rho^2\frac{\partial^2{E/A}}{\partial{\rho^2}}=322$~MeV)  
compared with that obtained using GM3 ($k^{-1}=240$~MeV). 

Observe that for $\rho \gtrsim 3 \rho_0$ the effect 
of the partial restoration of the chiral symmetry manifests as a softening of the EOS.
This softening is clearly visible in Fig.~\ref{bulk} 
where the bulk modulus computed with the CDM
equals the one evaluated using GM3 at about $3 \rho_0$ and becomes 
smaller for larger values of $\rho$. 
The effect of the isospin density is to increase the bulk modulus, since 
the $\rho$ meson
gives a positive contribution to the pressure proportional to 
the baryon density. 
The increase of the bulk modulus when moving from
Z/A=0.5 to Z/A=0.2 is of about 10\% both for the CDM and for GM3.

The dependence of the bulk modulus on the density and the temperature
can be measured in HIC experiments, in particular by analyzing the so
called collective flows.  A first attempt has been done in
Ref.~\cite{Russkikh:2006ae}, where they have extracted the value of
the incompressibility parameter from HIC experiments, observing a
significant softening of the EOS when the beam energy varies from 2 A
GeV up to 10 A GeV. By comparing with the CDM one can notice that this
softening can be interpreted as due to partial chiral symmetry
restoration if densities of at least~$6 \-- 7 \rho_0$ are reached
during a HIC. In Ref.\cite{Bonanno:2007kh} we have shown that if such
large densities are not reached, then this softening can better be
interpreted as due to the formation of a mixed hadron-quark phase.

\subsection{Effective meson masses}
A very important point concerns the predictions of the CDM 
about the experimentally measurable ``in medium'' masses, as the 
vector meson masses and the pion mass, whose expressions are provided
in Eqs.~(\ref{massaomega},\ref{massarho},\ref{massapione}).
\subsubsection{Effective pion mass}

\begin{figure}[ct!]
\centering
\includegraphics[scale=0.35]{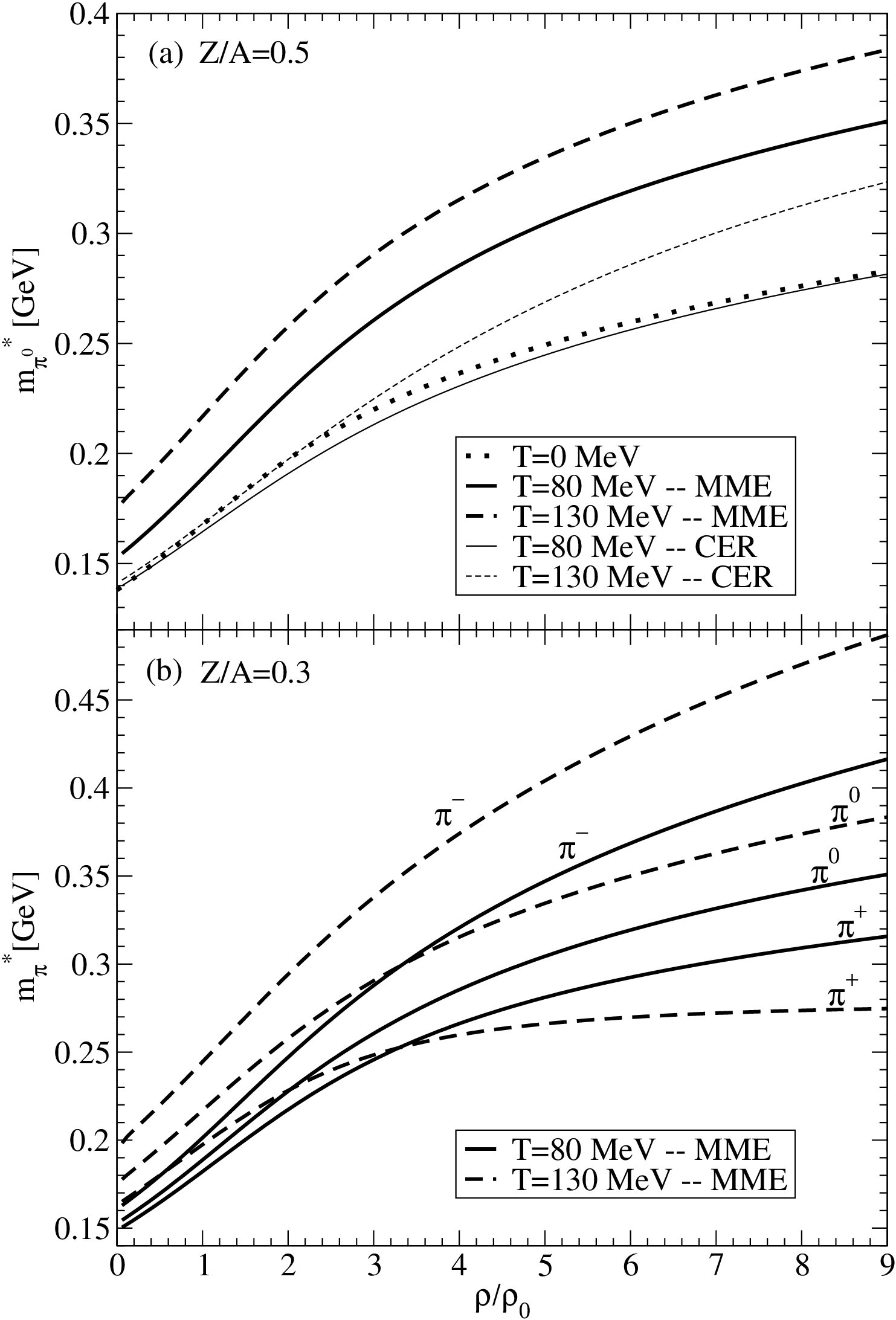}
\caption{\label{mpi}
\footnotesize In panel (a) the pion effective mass
obtained using the technique 
of Ref.~\cite{Mocsy:2004ab} (MME, thick lines)
is compared with the mass obtained using
Refs.~\cite{Carter:1996rf,Carter:1997fn} (CER, thin lines). 
The mass is plotted as a 
function of the baryon density,
for Z/A=0.5 and for three values of the temperature. The two techniques are 
equivalent at $T=0$.
In panel (b) we show the masses of the different isospin components of the pion, 
as a function of the baryon density, 
for Z/A=0.3 and for two values of the temperature. Here we adopted
the technique of Ref.~\cite{Mocsy:2004ab}.
At $T=0$ no isospin mass splitting is present.}
\end{figure}

The binding energy of the deeply bound 1s and 2p states in pionic
atoms of $^{207}Pb$, established experimentally through the $^{207}Pb(d,^{3}He)$ reaction, 
has been used 
to derive the value of the pion mass at nuclear matter 
density \cite{Geissel:2001px,Friedman:1998ed}.
The result of this still preliminary analysis is that
the pion mass increases by about 28 MeV at $\rho_0$
in symmetric nuclear matter. 

In Fig.~\ref{mpi}a the pion effective mass in symmetric 
nuclear matter is plotted as a function 
of the baryon density and for different values of the temperature.
We observe that the pion mass increases monotonously with the baryon 
density reaching a value of about 
167 MeV at $\rho_0$ and $T=0$, a result very close to the one suggested in 
Refs.\cite{Geissel:2001px,Friedman:1998ed}.
We also compare the technique of Refs.~\cite{Carter:1996rf,Carter:1997fn}
with the technique
of Ref.~\cite{Mocsy:2004ab} 
(see the previous discussion in subsection~\ref{thermal_flu}). Most of the results obtained in the present work
are almost independent of which technique is adopted, but the masses of the chiral fields
are very delicate and the two techniques provide different results even in the case of 
isospin symmetric matter, as we show in Fig.~\ref{mpi}a\\
In Fig.~\ref{mpi}b we show the effect of the isospin 
asymmetry (Z/A=0.3) on the pion mass. 
Notice that in isospin asymmetric matter the positive and negative pions have opposite (nonzero)
chemical potentials. Then, according to Eq.~(\ref{thep}), their
thermal fluctuations are different as well as their 
effective masses, defined by Eq.~(\ref{massapione}). In order to consider this 
effect, in our calculations we used the technique of Ref.~\cite{Mocsy:2004ab}.
In Fig.~\ref{mpi}b we show that the mass splitting of the different pion components 
increases substantially with the density and with the temperature (the mass splitting vanishes at T=0).
In particular, at $\rho=3\rho_0$ and T=130 MeV, the difference between $m_{\pi^-}$ and 
$m_{\pi^+}$ reaches a value of about 100 MeV. Notice that this result, obtained without
including any contribution associated with the so-called particle-hole excitations, is
comparable to the isospin splitting of the pion mass associated with the so-called Tomozawa-Weinberg term.

\subsubsection{In medium vector meson masses}

\begin{figure}[ct!]
\centering
\includegraphics[scale=0.35]{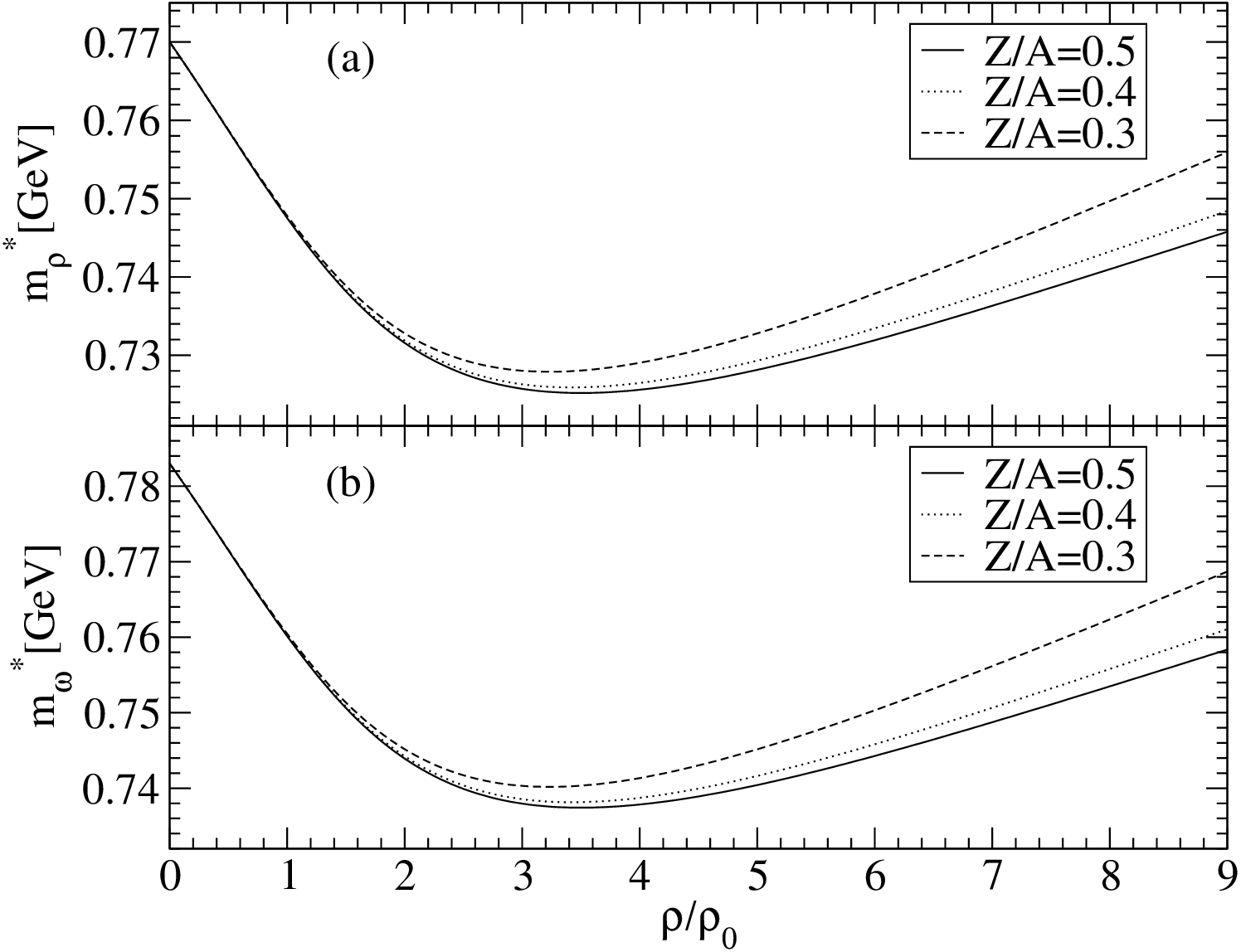}
\caption{\label{mvec}
\footnotesize Vector meson masses as a function of the baryon density at T=0 
and for different proton fractions.}
\end{figure}

The experimental determination of the in-medium modification 
of the light vector mesons ($\omega$, $\rho$ and $\Phi$) is still uncertain. 
Recent experiments have advanced to a level that in-medium spectral
information on vector mesons can be deduced. 
Experimentally, the best
approach is the dilepton spectroscopy, in particular
the investigation of the dileptonic decay modes of the light vector mesons,
produced off nuclei or in HICs.
In Fig.~\ref{mvec} the masses of the vector mesons $\omega$ and $\rho$, 
evaluated within the CDM, are plotted 
as functions of the baryon density at zero temperature, and for different 
values of the proton ratio. 
We do not show the values of the vector mesons masses for 
temperatures different from zero because, up to $T\sim 150$~MeV, the masses 
vary only slightly with the temperature (the differences are of a few per cent). 

Notice the peculiar behavior of 
$m_{\omega}$ and $m_{\rho}$ with $\rho$: 
for low densities the masses initially drop, reaching a
minimum in the neighbor of $\rho=3.5\rho_0$, while for
higher values of $\rho$, the masses  
increase again. This effect is due to the difficulty in restoring the scale invariance
at large densities, as discussed in subsection~\ref{scaleres}. 
The special scenario obtained within the CDM seems to be 
in agreement with several experimental analysis indicating a moderate 
reduction of the masses of the vector mesons at densities of the order of 
$\rho_0$ and no significant reduction when the masses are tested at 
very large densities.

For instance, in experiments of scattering of photons and protons on nuclei,
densities about $\rho_0$
are tested and the results indicate a drop
of the $\omega$-mass by 14\% at saturation density (CBELSA/TAPS) \cite{Trnka:2005ey}. More 
controversial is the situation concerning the $\rho$-mass:
the KEK-PS E325 collaboration indicates a drop by 9\% at $\rho_0$ 
\cite{Muto:2005za,Metag:2007zz} while a recent experiment at JLAB suggests a 
smaller drop by $\sim$~2\% at $\rho_0$ \cite{:2007mga}.   
On the other hand, in HIC experiments at energies of the order
of one hundred A~GeV (SPS and NA60), the vector mesons mass shift has been found
to vanish \cite{Adamova:2002kf,Arnaldi:2006jq,Metag:2007zz}. This result presumably 
discards the scenario 
predicted by Brown and Rho \cite{vanHees:2006iv,Metag:2007zz}, where 
the vector meson masses drop continuously with the density \cite{Brown:1991kk}.

Finally, the effect of a finite isospin density is to increase the vector masses as it can be 
observed in Fig.~\ref{mvec}. This effect
grows with the baryon density, although at $\rho=9\rho_0$ the increment is only
of about 1\% for both the vector meson masses.

\section{Astrophysical applications}
\label{astro}

Neutron Stars (NSs) are the most dense objects in nature while their 
temperature is always below a few ten MeV and typically lower than one MeV. 
Therefore they are the only objects which can provide 
information on the EOS of matter at very large densities and low temperature. 
Unfortunately the structure of compact stars is still uncertain since 
astrophysical data can provide only rather indirect indications, leaving 
the field open to
theoretical investigations based on models for the high density EOS.
Since the CDM is a chiral model, it is very interesting
to investigate its previsions concerning matter 
at very large densities
and small temperatures. In particular it is possible to
explore how the chiral dynamics can affect 
the structure and the formation of a NS. 
The dynamics of the gluon 
condensate does not play a crucial role at low temperatures.
In this section we compute the value of the adiabatic index 
which plays a crucial role in 
Supernova explosions and we investigate within the CDM 
the structure of NSs \cite{Dexheimer:2008ax}. 

In order to better understand the 
formation and the structure of a NS
we also discuss the effective masses of the various fields
and, finally, we consider the possibility 
to reach the pion condensation in NSs.

\subsection{Adiabatic index}
As far as we know, the explosion of a core-collapse
supernova proceeds as the following
(for a recent review see Ref.~\cite{Woosley:2006ie}): when a massive star exhausts 
its nuclear fuel, the gravitational
force cannot be contrasted by the outward pressure and the core starts
to collapse. During this stage, which typically last much less than a
second, the pressure strongly exceeds the electron degeneration
pressure and neutronization occurs. The temperature is of the
order of few ten MeV and the neutrinos cannot escape from the collapsing matter
since their cross section is very large. The lepton number is then
conserved and no energy can escape the system, giving the possibility to
describe the entire process as an adiabatic compression with a
entropy per baryon of the order of unity. Moreover, since the time
scale of the $\beta$ interactions is of the order of $10^{-8}$ s, the system has
the time to reach the $\beta$-equilibrium. When the density of the
core overcomes $\rho_0$, due to the
nuclear repulsion the core bounces triggering a shock wave which
propagates to the outer layers of the star. In 
hydrodynamical simulations, it turns out that the shock wave stalls
while it is still propagating inside the 
outer layers of the progenitor star. Therefore the so called 
``direct mechanism'', characterized by
the idea that the explosion is directly related to the shock wave
generated by the bounce, has been ruled out. The failure of this
direct mechanism sets the stage for a `delayed'
mechanism, in which the shock is re-energized by the heating of 
the outer layers due to the intense neutrino flux emerging from the
neutrino-sphere.

It is worth discussing more in details the conditions under which
the direct mechanism would be successful. It is has been noticed that 
the possibility for a supernova to explode
via the ``direct mechanism'' is related to
the softening of the EOS at densities just above $\rho_0$ \cite{SLM}. This possibility,
exploited in the Baron-Cooperstein-Kahana (BCK) EOS \cite{Baron:1985gg} has been discarded because
it seems to be inconsistent with the constraints on the NS masses
coming from observations, since a too soft EOS cannot
support a sufficiently massive NS. 
A more sophisticated approach invokes a softening taking place only 
in a finite density window and associated with a phase transition.
In this way the EOS can be stiff again at densities larger than the ones
associated with the transition. This possibility is also not new and it has been investigated
in several papers 
\cite{Migdal:1979je,Takahara:1985,Gentile:1993ma,Drago:1998qu}.
In all those works a strong
softening of the EOS is ascribed to a phase transition 
which can be due either to the pion condensation or to the formation of 
a mixed phase of hadrons and quarks. Since in the CDM we obtain a softening of the
EOS at moderate densities due to chiral symmetry restoration it is worth to explore
the possible influence of this transition on the supernova explosion.

In order to investigate the core collapse of a massive star,
we compute the EOS along adiabatic paths for 
matter composed of neutrons, protons, electrons and electronic neutrinos.
Due to the previous considerations, the conditions imposed to the system are:  
\begin{flushleft}
\textmd{i) leptonic number conservation:}
\end{flushleft}
\bq
\rho_e+\rho_{\nu_e}=Y_{l_e}(\rho_p+\rho_n)
\eq
\begin{flushleft}
\textmd{ii) electric charge neutrality:}
\end{flushleft}
\bq
\rho_p=\rho_e
\eq
\begin{flushleft}
\textmd{iii) $\beta$-equilibrium:}
\end{flushleft}
\bq
\mu_n-\mu_p=\mu_e-\mu_{\nu_e}
\eq
where $Y_{l_e}=0.4$ is the constant ratio of the leptonic density to
the baryon density. Since before the collapse no muons are present, we keep
$Y_{l_{\mu}}=\frac{\rho_{\mu}+\rho_{\nu_{\mu}}}{\rho}=0$.

\begin{figure}[t]
\includegraphics*[scale=0.35]{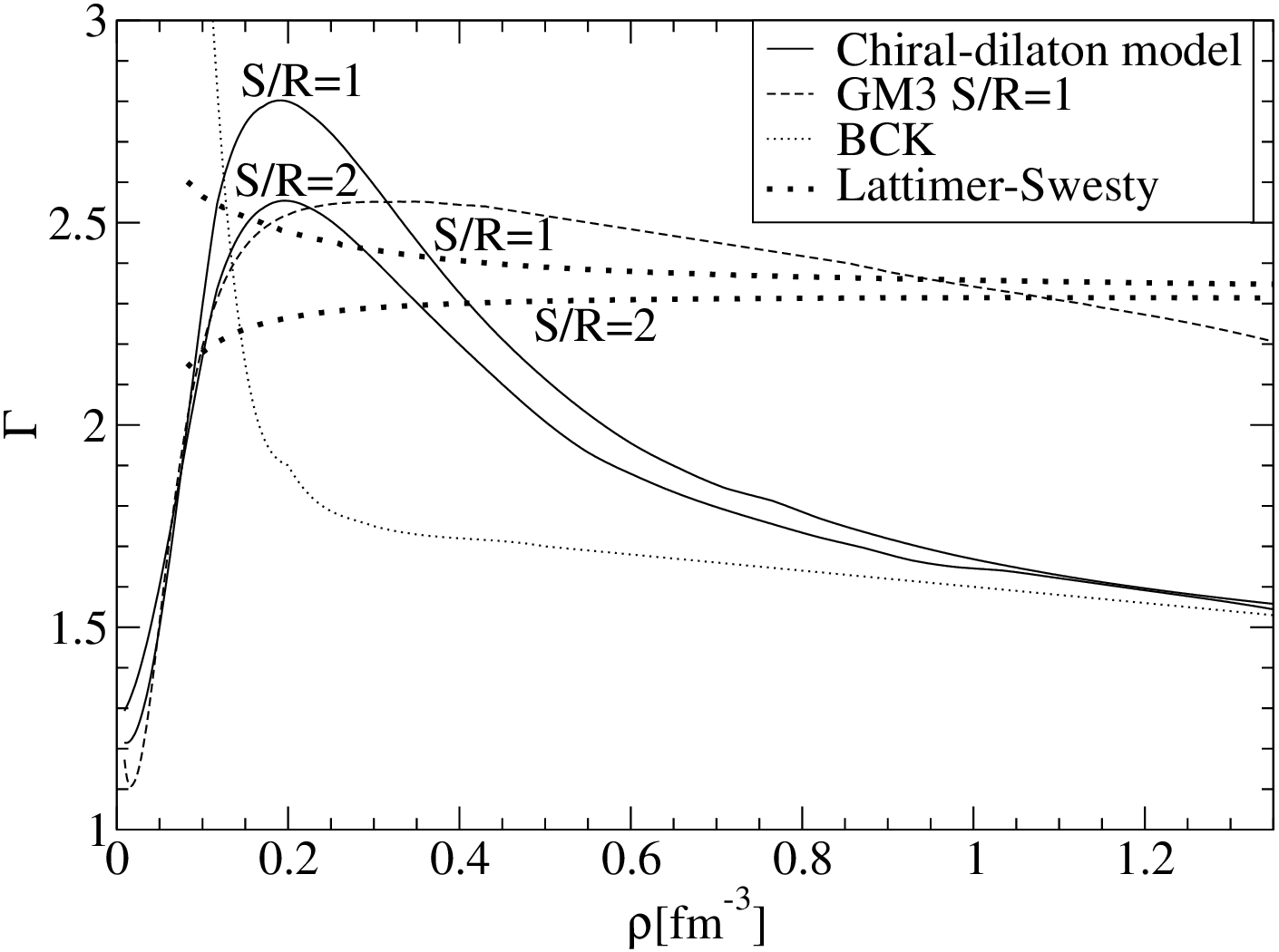}
\caption{\label{adiab}
\footnotesize Adiabatic index as a function of $\rho$ 
computed within various models for isoentropic $n,p,e^-,\nu_{e}$ matter with
$Y_{l_e}=0.4$.}
\end{figure}

\begin{figure}[t]
\centering
\includegraphics*[scale=0.35]{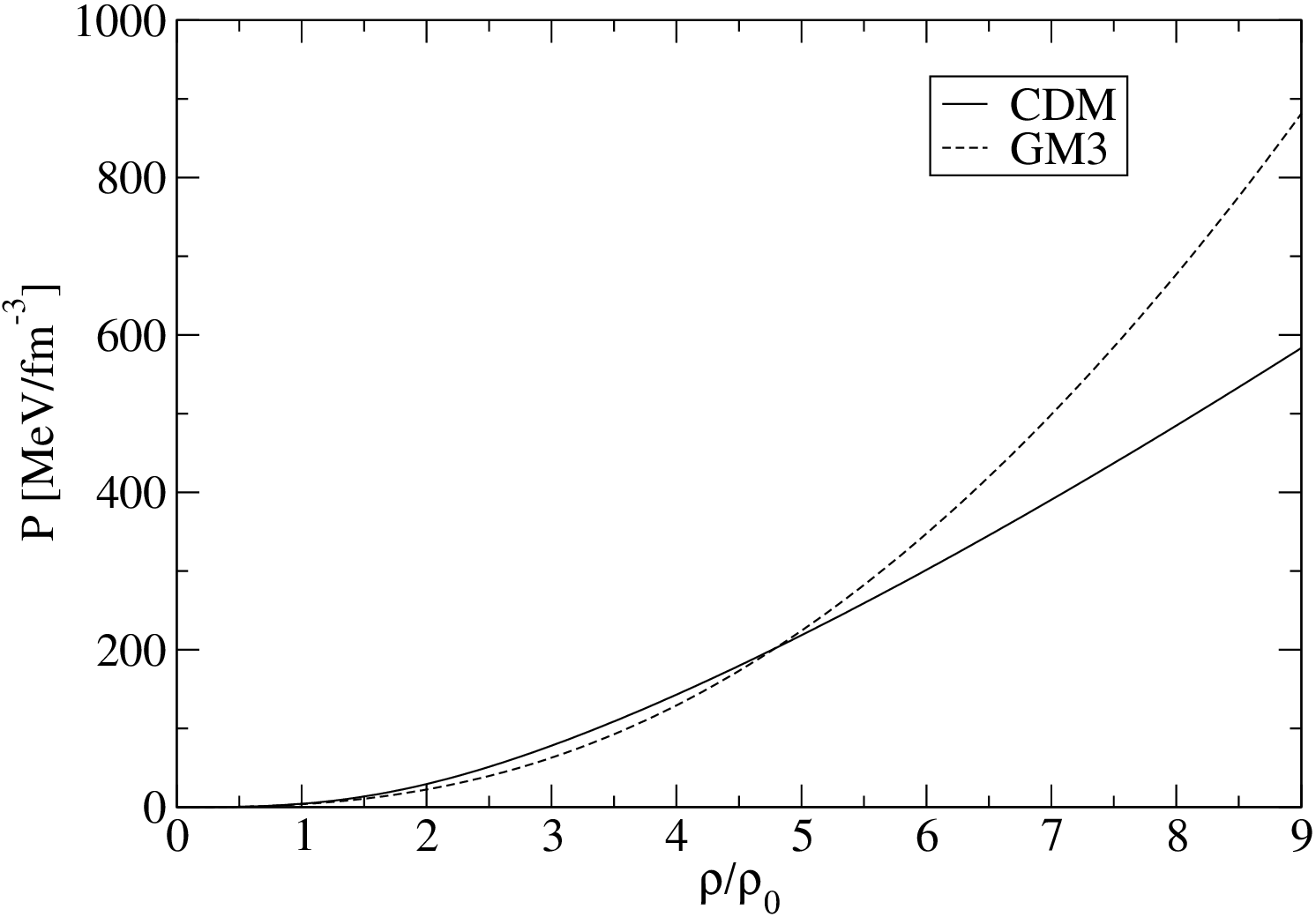}
\caption{\label{pb}
\footnotesize Pressure for NS matter as a function of baryon density, 
evaluated within the CDM and within the Walecka GM3 model.}
\end{figure}

An important quantity which allows to estimate the softness of the EOS
is the adiabatic index, which is defined as:
\begin{equation}
\Gamma=\frac{\partial \ln P}{\partial \ln\rho} 
\end{equation}
where P is the pressure. It 
corresponds to the index of a polytropic equation of state
$P(\rho)=P_0\left(\frac{\rho}{\rho_0}\right)^{\Gamma}$.
In Fig.~\ref{adiab} we show $\Gamma$, obtained from the CDM EOS and from other 
models, as a function of $\rho$. 
Notice that within the Walecka GM3 and the Lattimer-Swesty 
EOS \cite{Lattimer:1991nc} no significant reduction of $\Gamma$ is 
observed at supranuclear densities. 
At the contrary, the strong softening occurring within the BCK 
EOS \cite{Baron:1985gg} at densities just above $\rho_0$ 
allows a supernova to explode via the ``direct mechanism'' (recall that 
the BCK EOS has been ruled out since it provides 
too low NS masses).
Within the CDM the partial restoration of chiral symmetry 
manifests as a strong softening of the EOS (see Fig.~\ref{pb}). 
This softening is also observed as a strong reduction of $\Gamma$ 
for densities larger than $3\rho_0$. 
Although this reduction is comparable with that obtained in BCK, 
since the density at which it occurs is much 
larger than $\rho_0$ we do not expect that the CDM can allow a supernova 
to explode via the ``direct mechanism''.
On the other hand, since very large densities are reached 
during the collapse of a binary system of neutron stars, we expect that such a 
strong decrease of $\Gamma$ can have interesting consequences, 
for instance in affecting the gravitational wave signal emitted just before 
the formation of the black-hole~\cite{janka}.

\subsection{Neutron stars} 

A NS is the cold remnant of the explosion of a core-collapse supernova 
and it is composed by neutrons, protons, electrons and muons in $\beta$-equilibrium.
Since the temperature is very low the neutrinos cross section is very small, 
allowing the neutrinos produced by $\beta$-decay to
escape from the star. Therefore the lepton number is not conserved.
Since electric charge neutrality has also to be imposed, 
the conditions defining NS matter read:

\bqr
\mu_{n}=\mu_p+\mu_e\nonumber\\
\mu_{n}=\mu_p+\mu_{\mu}\nonumber\\
\rho_{p}=\rho_e+\rho_{\mu}\, .
\eqr
That EOS is then used to solve the Tolman-Oppenheimer-Volkov 
(TOV) equation obtaining a sequence of stable stellar configuration.

\begin{figure}[t]
\includegraphics[scale=0.35]{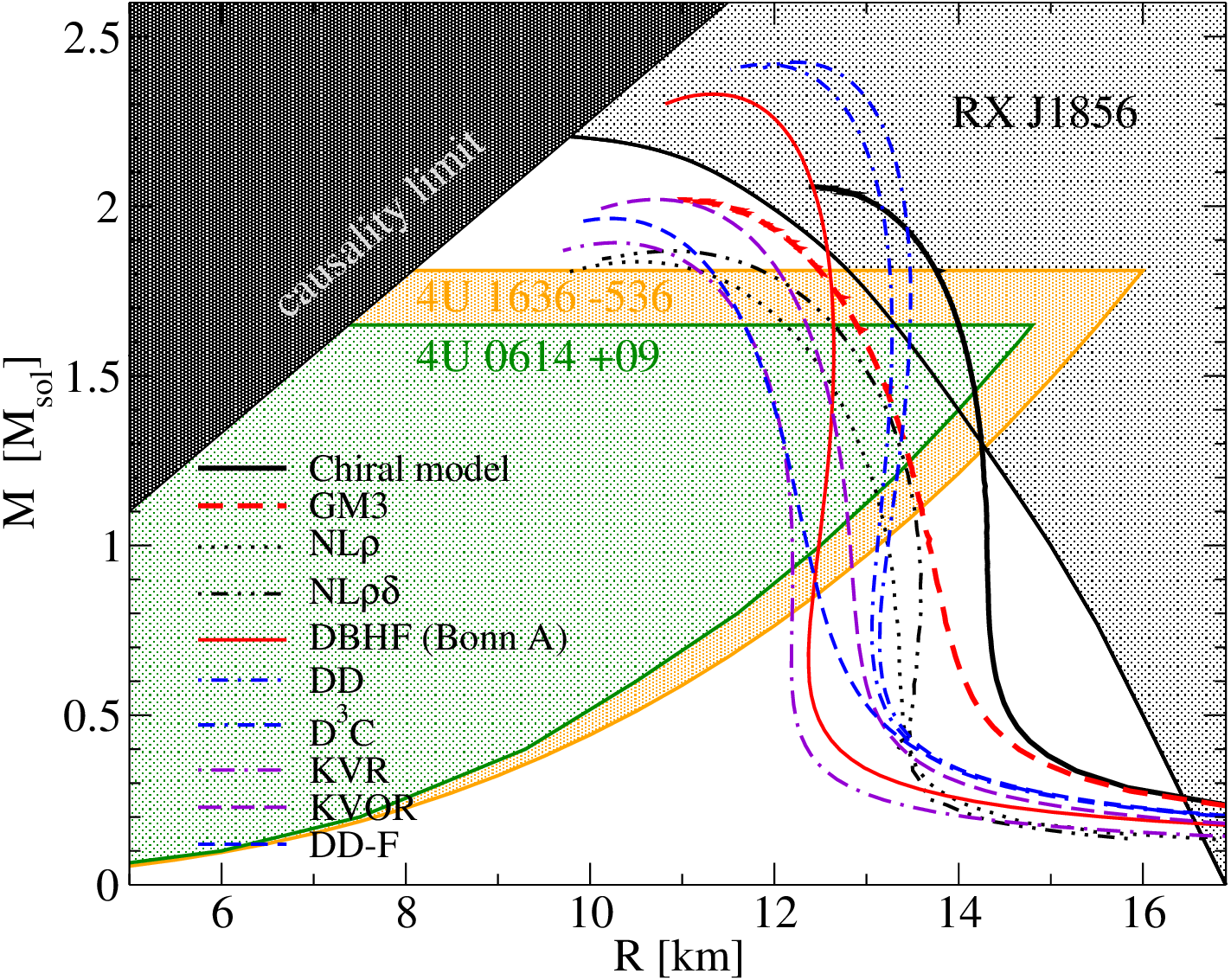}
\caption{\label{mr}
\footnotesize (Color online) Mass-radius relation for the CDM and 
for various other models together with recent experimental constraints.
Adapted from Ref.~\cite{Klahn:2006ir}.}
\end{figure}
The mass-radius relations for the CDM and various other models 
are shown in Fig.~\ref{mr}, together with several recent experimental 
constraints coming from observations and discussed in Ref.~\cite{Klahn:2006ir}. 
The models considered in Fig.~\ref{mr} can be separated into Relativistic Mean Field (RMF) 
phenomenological models, including the Walecka type models (GM3 
\cite{Glendenning:1991es}, NL$\rho$ 
\cite{Gaitanos:2003zg}
and NL$\rho\delta$ \cite{Liu:2001iz}) and those with density dependent coupling 
constants (DD \cite{Typel:1999yq}, $D^3$C and DD-F 
\cite{Typel:2005ba,Klahn:2006ir}), the 
``ab-initio'' calculations including the relativistic 
Dirac-Brueckner-Hartree-Fock (DBHF 
\cite{vanDalen:2004pn,vanDalen:2005ns,GrossBoelting:1998jg,deJong:1997hr}) 
and the hybrid models 
(KVR and KVOR \cite{Kolomeitsev:2004ff,Akmal:1998cf}) 
which bridge the gap between the fully microscopic and more phenomenological
descriptions. 

The experimental constraints are extracted from observations 
of kilohertz quasi-periodic brightness oscillations in the Low 
Mass X-Binaries 4U 0614+09 
and 4U 1636-536 and from the thermal radiation of the isolated NS source 
RX J1856.5-3754.

Due to the rather unrealistically large value of the incompressibility 
parameter at $\rho_0$ ($K^{-1}$=322 MeV), larger 
NS radii are obtained within the CDM than within the other models.
Therefore, although the CDM allows to have a NS
with a mass $M\sim 1.4 M_\odot$ and a radius of 14~Km, satisfying the 
constraints from RX J1856 (the other models 
exclude this interpretation), our result is questionable due to the too large
value of the incompressibility at saturation. 

\begin{figure}[t!]
\includegraphics[scale=0.35]{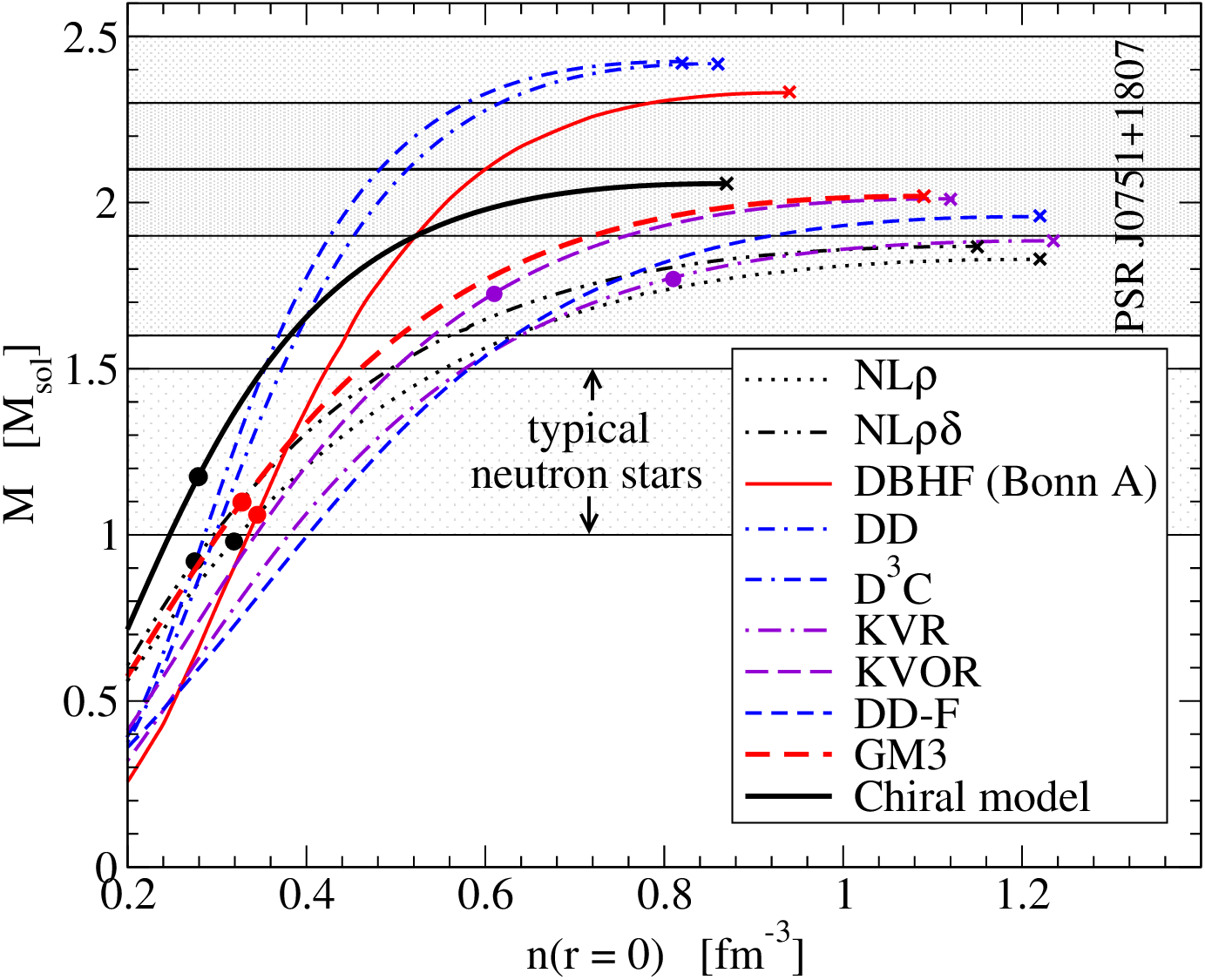}
\caption{\label{mn}
\footnotesize (Color online) NS mass vs. central density for the same models of
Fig.~\ref{mr}. The full 
circles indicate the DU thresholds and the stars indicate the maximum mass
configurations. Adapted from Ref.~\cite{Klahn:2006ir}.}
\end{figure}

An important check is based on the recent observations
of PSR J0751+1807. The data analysis, although still uncertain, allows 
to extract a NS mass of 
$2.1\pm0.2\left(^{+0.4}_{-0.5}\right) {\rm M_\odot}$ 
(first error estimate with $1\sigma$ confidence, second in brackets with
 $2\sigma$ confidence) \cite{Nice:2005fi}.
Since the CDM exhibits a maximum mass $M=2.06M_\odot$ 
this constraint is easily fulfilled. 
The constraint is displayed in Fig.~\ref{mn} where the NS mass vs. central 
density relation is shown for the same models of Fig.~\ref{mr} and where 
the crosses denote the maximum mass configurations. 

When discussing NSs a very important role is played by the cooling mechanism.
We can roughly divide the cooling mechanisms in ``slow'' ones, as e.g. 
the so-called modified Urca process, and fast ones, as e.g. 
the Direct Urca (DU) process which corresponds to a sequence of 
neutron $\beta$-decays and 
lepton captures (see e.g. Ref.~(\cite{Haensel:2007yy})). 
From simple considerations it is possible to find the critical baryon density 
beyond which the DU processes can occur in a NS, which turns out to be the 
density at which the proton ratio in NS matter reaches the value:
\bq
x_{DU}=\frac{1}{1+(1+x_e^{1/3})^3}
\eq
where $x_e=\frac{\rho_e}{\rho_e+\rho_{\mu}}$.   
In Fig.~\ref{mn} the DU-thresholds are denoted by full circles. 
Notice that, although the CDM allows the DU to occur for
NS masses exceeding $\sim 1.2 M_\odot$, its threshold is slightly
higher than those obtained within GM3, NL$\rho$, NL$\rho\delta$ and DBHF.

Finally, the effect of the softening of the CDM EOS manifests as an 
``anomalous'' behavior of the NS sequence with respect to the other models.
Notice in fact that in Fig.~\ref{mn} at low central densities 
the CDM NS masses are rather large and the CDM sequence seems to follow the 
trend of the $DD$, $D^3C$ and $DBHF$ 
models whose maximum mass configurations exceed 2.3$M_\odot$. 
However at larger central densities the sequence flattens and a maximum mass 
only slightly larger than GM3 is reached ($M_{max}=2.06 M_\odot$).

\subsection{In medium masses}
\begin{figure}[t!]
\centering
\includegraphics[scale=0.35]{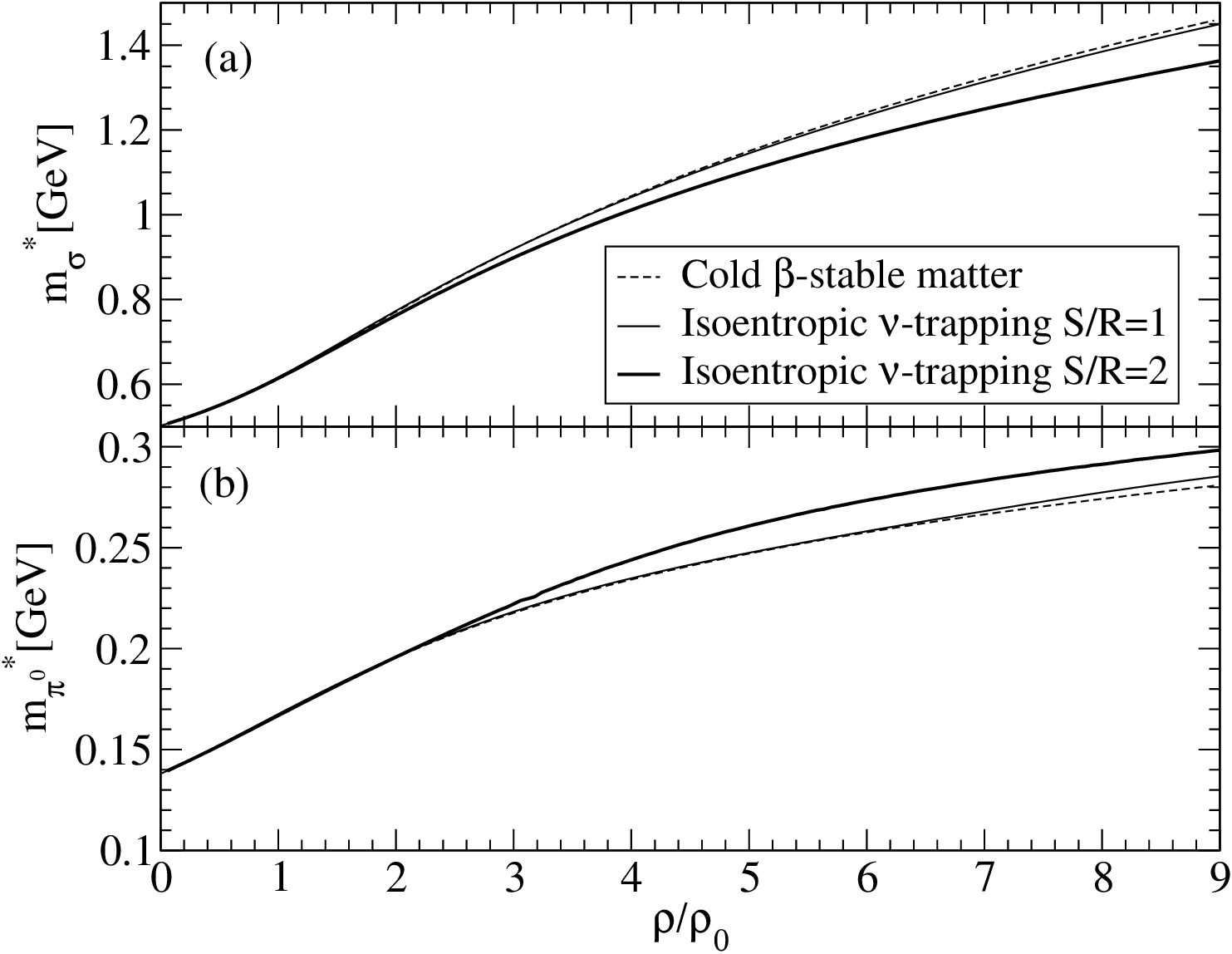}
\caption{\label{np2}
\footnotesize $\sigma$ (figure (a)) and $\pi^0$ (figure (b)) effective masses as a function of 
the baryon density for NS matter (dashed line) 
and for isoentropic $\nu$-trapped matter with entropy per nucleon S/R=1 
(thin solid line) and S/R=2 (bold solid line).}
\end{figure}

In order to understand how the partial restoration of chiral symmetry 
affects the structure of NSs, we computed, within the CDM and 
using the technique developed in Ref.~\cite{Mocsy:2004ab}, the 
$\pi^0$ and $\sigma$ 
meson effective masses as a function of $\rho$ both
for NS matter and for 
isoentropic neutrino-trapped matter (see Fig.~\ref{np2}).
We find that even at large densities the masses of the two chiral fields 
are still rather different, although the mass of 
$\pi^0$ is considerably larger than in the vacuum.
The effect of the temperature on the partial restoration of chiral 
symmetry is very small but it can anyway be noticed by comparing the 
isoentropic curves with S/R=1 and S/R=2: for the larger value of the 
entropy the $\sigma$ and $\pi^0$ masses start getting closer,
the $\sigma$ mass by decreasing and the $\pi^0$ mass by increasing its value.

\begin{figure}[t!]
\centering
\includegraphics[scale=0.35]{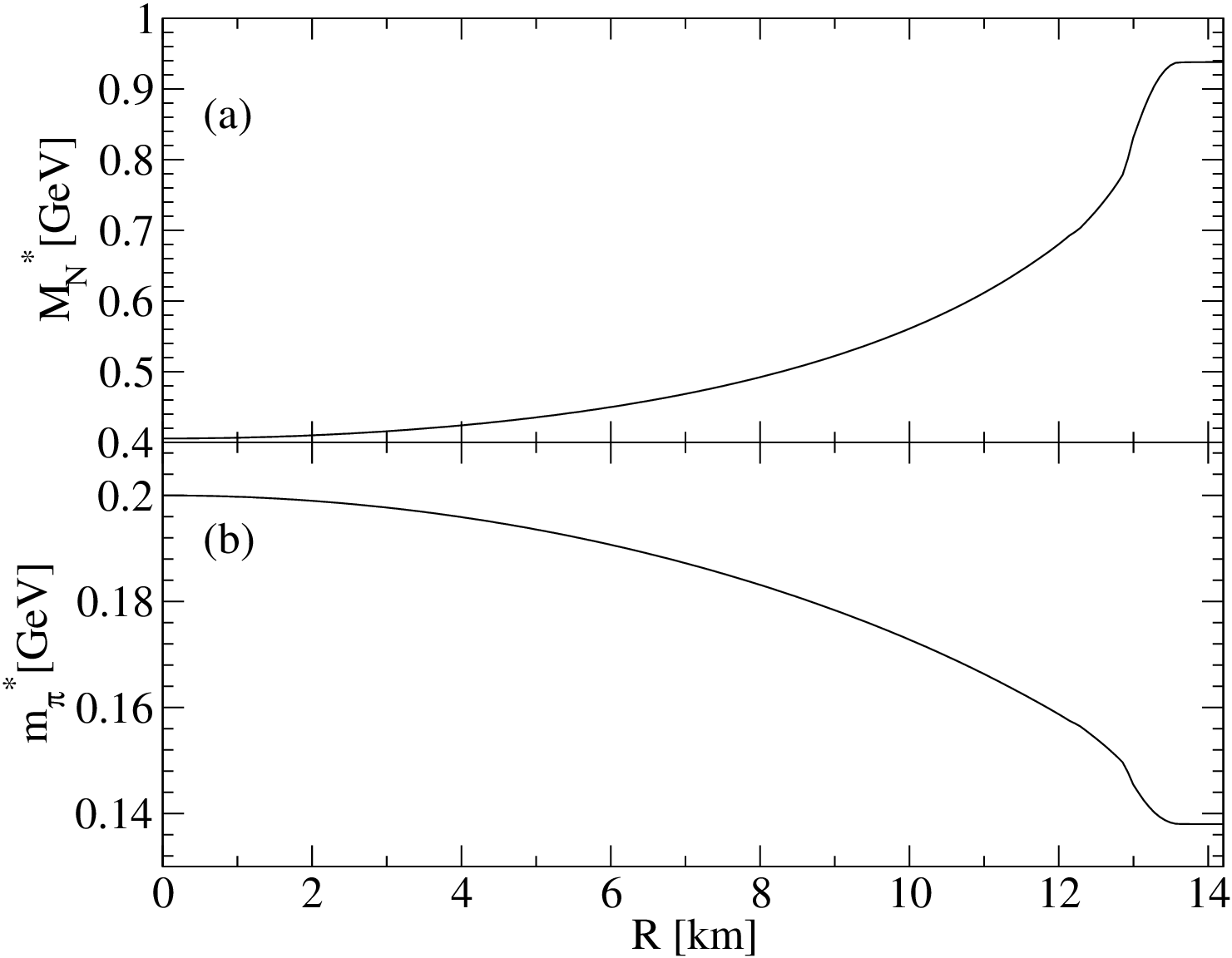}
\caption{\label{np}
\footnotesize In-medium mass modification of the nucleon~(figure (a)) 
and of the pion~(figure (b)) inside a neutron star of 1.4~M$_{\odot}$.}
\end{figure}

Finally, we have examined the in-medium modification of the nucleon and 
of the pion mass by 
plotting their values as a function of the radius in the case of
a spherical non-rotating NS of mass $M=1.4 M_{\odot}$, 
see Fig.~\ref{np} (notice that, although in presence of isospin asymmetry, at T=0 
it comes out that $m_{\pi^0}=m_{\pi^+}=m_{\pi^-}\equiv m_{\pv}$).
The value of nucleon mass 
decreases by more than $50\%$ when going from the outer crust 
to the center of the star. The pion mass, at the contrary, increases by about 
$45 \%$. 

\subsection{Pion condensate}
\label{bose}
Since in compact stars very large baryon densities and 
isospin asymmetries are reached, it is interesting to investigate 
the possibility to form a meson Bose-Einstein Condensate (BEC). 
In particular we investigate here the possibility of reaching the pion BEC, 
which is easier to achieve due to the small value of the pion mass.
Usually with ``pion condensate'' one refers
to a collective pionic excitation at non zero-momentum. This collective 
state is a pole in the pion propagator modified by the inclusion 
of the particle-hole excitations.
In this work we do not include the particle-hole excitations in order to
satisfy the Goldstone theorem at finite temperature (see the discussion 
in section \ref{goldstone}) 
and we call  ``pion condensate'' the zero-momentum Bose-Einstein
condensate of pions.

Recall that in order to 
conserve the isospin charge a chemical potential is assigned to 
the pions, which reads: 
\bq
\mu_{\pi^+}=\mu_p-\mu_n \, ,
\eq  
where $\mu_p$ and $\mu_n$ are the chemical potential of protons and neutrons, 
respectively. The equation above assures the equilibrium in the reaction 
$n\leftrightarrow p+\pi^-$.
Moreover the pion effective chemical potential, which enters the 
thermodynamical integrals, is given by:
\bq
\mu^*_{\pi^+}=\mu^*_p-\mu^*_n \, ,
\eq
where $\mu_p^*$ and $\mu_n^*$ are the effective chemical potential of protons 
and neutrons respectively, given by Eq.~(\ref{potchim}).
The effective chemical potential of the pion increases with 
the baryon density and with the isospin asymmetry.
When the effective chemical potential becomes equal to the pion effective mass, 
real pions start to be produced in the medium and they condense in the state 
of minimum energy, i.e. the zero momentum mode. 
The condition for the onset of the pion BEC then 
reads \cite{Kapusta:2006pm}:
\bq
\mu^*_{\pi^-}=m^*_{\pi} \, .
\eq

\begin{figure}[t!]
\includegraphics*[scale=0.35]{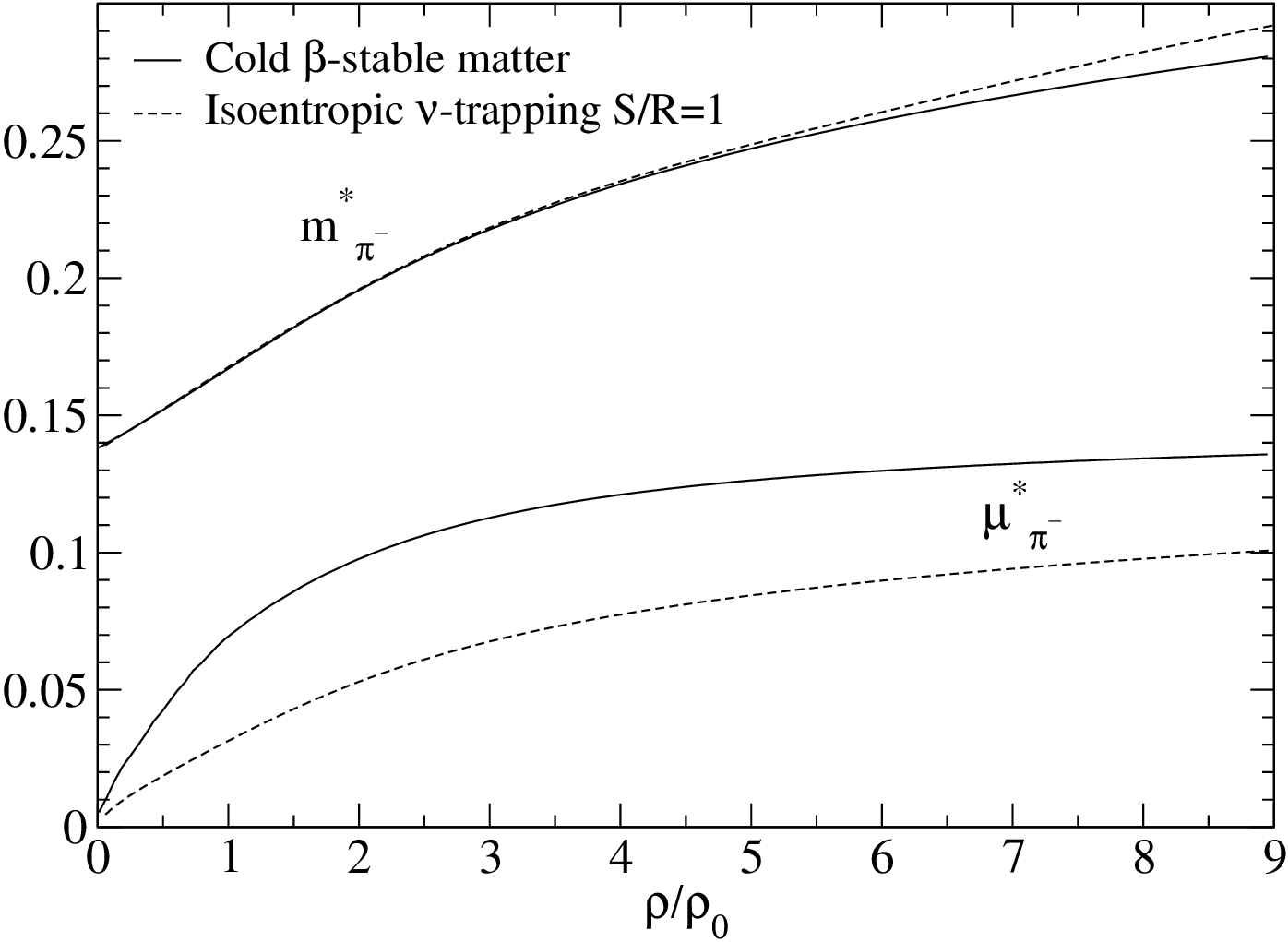}
\caption{\label{cond}
\footnotesize $\pi^-$ effective mass and effective chemical potential 
as a function of the baryon density, computed within the CDM 
for NS matter and for isoentropic $\nu$-trapped 
matter with entropy per baryon S/R=1.}
\end{figure}

\noindent In Fig.~\ref{cond} the $\pi^-$ effective mass 
and its effective 
chemical potential (computed using the technique developed in Ref.~\cite{Mocsy:2004ab})
are plotted as a function of $\rho$. Here we 
consider two different cases: $\beta$-stable NS matter 
at T=0 and  
isoentropic $\nu$-trapped matter at $\beta$-equilibrium with entropy per 
baryon S/R=1.
As it can be seen the $\pi^-$ mass grows with $\rho$ and it does not 
change appreciably between the two cases.
Also $\mu_{\pi^-}$ increases 
with $\rho$, but its value actually depends on the value of the 
entropy and on neutrino-trapping.

The important result is that the condition for the onset of the 
BEC is never achieved, not even at large densities. This is due to
the increase of the pion effective mass with $\rho$, 
which remains always larger than the chemical potential.
In simpler models, like e.g. the Walecka GM3 where the pion mass is 
constant, the increase of the chemical potential gives rise to 
the pion condensation, as discussed in Ref.~\cite{Muller:1997tm}.

\section{Conclusions}
\label{Conclusions}

The main aim of this work was to investigate the behavior
of matter at large densities and temperatures by using a effective
lagrangian with broken scale invariance and chiral symmetry.
Our results are concordant with those 
obtained in Refs.~\cite{Carter:1996rf,Carter:1997fn}. 
We extended the previous calculations in several directions and wherever
possible we compared with experimental data.
We have also used two different techniques to evaluate the 
effects of thermal fluctuations, i.e. the method developed in
Refs.~\cite{Carter:1996rf,Carter:1997fn} versus the one developed
in Ref.~\cite{Mocsy:2004ab}. The latter is more general since it does not
make any assumption concerning the value of the thermal fluctuations of
the chiral fields. The two techniques 
provide similar results as long as we are not investigating subtle isospin 
dependencies such as the mass splitting of the isospin components of the pion.

Our main results are the following:

\noindent -- we have 
provided a phase diagram mapping the regions in which chiral symmetry
and/or scale invariance are restored. The 
scenario provided by the model we have studied, i.e. the
decoupling of chiral symmetry
restoration and scale invariance restoration (which is likely associated
with deconfinement) shares strong similarities with the scenario recently suggested
by McLerran and Pisarski in the large $N_c$ limit \cite{McLerran:2007qj};

\noindent -- the masses of vector mesons are affected by the partial
restoration of scale invariance, since they scale with the 
dilaton field. Therefore 
scale invariance restoration is more difficult at large densities,
because the repulsive effect of the exchange of vector mesons is enhanced
if their mass is reduced;

\noindent -- the masses of the vector
mesons first reduce at finite density, but at larger densities
they even increase. This result holds up to temperatures of the order
of 200 MeV, at even larger temperatures scale invariance starts being restored and
the vector meson masses drop to zero;

\noindent -- In isospin symmetric matter the pion mass increases with the density 
and with the temperature. 
At $\rho_0$ and T=0 the effective mass of the pion is about 170 MeV.\\ 
In presence of isospin asymmetry the masses of the different pion components split.
The splitting increases with the density and the temperature and, 
for instance, at $\rho=3\rho_0$, T=130 MeV and Z/A=0.3 $m_{\pi^-}-m_{\pi^+}\sim 100$ MeV;

\noindent -- a very substantial reduction of the adiabatic index takes place
at densities exceeding 3 $\rho_0$. Although this softening 
cannot directly affect the mechanism of Supernova explosion,
because it starts at too large densities, its signature
could be observed during a two neutron star merging;

\noindent -- the BEC of pions does not form in NSs 
due to the increase of the pion mass with the density.

The present work can be extended in several directions. In particular
the restoration of scale invariance will be better investigated
by including in the calculation also the thermal contribution of 
the dilaton field, while in the present work we have only taken into account
the mean field effect.
Hyperons should also be included in the model. This direction is now explored
by analysis similar to ours and performed by the Frankfurt group
\cite{Dexheimer:2008ax}.
The most interesting extension would be to take into account
quark degrees of freedom. An attempt in that direction is a
recent paper in which deconfinement is discussed in a dilaton model,
but in the pure gauge sector~\cite{Drago:2001gd}.

\section{Acknowledgments}
It is a pleasure to thank Veronica Dexhaimer, J\"urgen Schaffner-Bielich and
Raffaele Tripiccione for many useful discussions,
David Blaschke and Thomas Klahn for providing data used in  many figures.
We are particularly grateful to Andrea Lavagno, with whom most of the 
problems investigated in the work have been extensively analyzed.

\bibliography{biblio}

\begin{thebibliography}{69}
\expandafter\ifx\csname natexlab\endcsname\relax\def\natexlab#1{#1}\fi
\expandafter\ifx\csname bibnamefont\endcsname\relax
  \def\bibnamefont#1{#1}\fi
\expandafter\ifx\csname bibfnamefont\endcsname\relax
  \def\bibfnamefont#1{#1}\fi
\expandafter\ifx\csname citenamefont\endcsname\relax
  \def\citenamefont#1{#1}\fi
\expandafter\ifx\csname url\endcsname\relax
  \def\url#1{\texttt{#1}}\fi
\expandafter\ifx\csname urlprefix\endcsname\relax\def\urlprefix{URL }\fi
\providecommand{\bibinfo}[2]{#2}
\providecommand{\eprint}[2][]{\url{#2}}

\bibitem[{\citenamefont{Heide et~al.}(1994)\citenamefont{Heide, Rudaz, and
  Ellis}}]{Heide:1993yz}
\bibinfo{author}{\bibfnamefont{E.~K.} \bibnamefont{Heide}},
  \bibinfo{author}{\bibfnamefont{S.}~\bibnamefont{Rudaz}}, \bibnamefont{and}
  \bibinfo{author}{\bibfnamefont{P.~J.} \bibnamefont{Ellis}},
  \bibinfo{journal}{Nucl. Phys.} \textbf{\bibinfo{volume}{A571}},
  \bibinfo{pages}{713} (\bibinfo{year}{1994}).

\bibitem[{\citenamefont{Carter et~al.}(1996)\citenamefont{Carter, Ellis, and
  Rudaz}}]{Carter:1995zi}
\bibinfo{author}{\bibfnamefont{G.~W.} \bibnamefont{Carter}},
  \bibinfo{author}{\bibfnamefont{P.~J.} \bibnamefont{Ellis}}, \bibnamefont{and}
  \bibinfo{author}{\bibfnamefont{S.}~\bibnamefont{Rudaz}},
  \bibinfo{journal}{Nucl. Phys.} \textbf{\bibinfo{volume}{A603}},
  \bibinfo{pages}{367} (\bibinfo{year}{1996}).

\bibitem[{\citenamefont{Carter et~al.}(1997)\citenamefont{Carter, Ellis, and
  Rudaz}}]{Carter:1996rf}
\bibinfo{author}{\bibfnamefont{G.~W.} \bibnamefont{Carter}},
  \bibinfo{author}{\bibfnamefont{P.~J.} \bibnamefont{Ellis}}, \bibnamefont{and}
  \bibinfo{author}{\bibfnamefont{S.}~\bibnamefont{Rudaz}},
  \bibinfo{journal}{Nucl. Phys.} \textbf{\bibinfo{volume}{A618}},
  \bibinfo{pages}{317} (\bibinfo{year}{1997}).

\bibitem[{\citenamefont{Carter and Ellis}(1998)}]{Carter:1997fn}
\bibinfo{author}{\bibfnamefont{G.~W.} \bibnamefont{Carter}} \bibnamefont{and}
  \bibinfo{author}{\bibfnamefont{P.~J.} \bibnamefont{Ellis}},
  \bibinfo{journal}{Nucl. Phys.} \textbf{\bibinfo{volume}{A628}},
  \bibinfo{pages}{325} (\bibinfo{year}{1998}).

\bibitem[{\citenamefont{Senger}(2004)}]{Senger:2004jw}
\bibinfo{author}{\bibfnamefont{P.}~\bibnamefont{Senger}}, \bibinfo{journal}{J.
  Phys.} \textbf{\bibinfo{volume}{G30}}, \bibinfo{pages}{S1087}
  (\bibinfo{year}{2004}).

\bibitem[{\citenamefont{Furnstahl et~al.}(1996)\citenamefont{Furnstahl, Serot,
  and Tang}}]{Furnstahl:1995zb}
\bibinfo{author}{\bibfnamefont{R.~J.} \bibnamefont{Furnstahl}},
  \bibinfo{author}{\bibfnamefont{B.~D.} \bibnamefont{Serot}}, \bibnamefont{and}
  \bibinfo{author}{\bibfnamefont{H.-B.} \bibnamefont{Tang}},
  \bibinfo{journal}{Nucl. Phys.} \textbf{\bibinfo{volume}{A598}},
  \bibinfo{pages}{539} (\bibinfo{year}{1996}).

\bibitem[{\citenamefont{Dexheimer
  et~al.}(2008{\natexlab{a}})\citenamefont{Dexheimer, Schramm, and
  Zschiesche}}]{Dexheimer:2007tn}
\bibinfo{author}{\bibfnamefont{V.}~\bibnamefont{Dexheimer}},
  \bibinfo{author}{\bibfnamefont{S.}~\bibnamefont{Schramm}}, \bibnamefont{and}
  \bibinfo{author}{\bibfnamefont{D.}~\bibnamefont{Zschiesche}},
  \bibinfo{journal}{Phys. Rev.} \textbf{\bibinfo{volume}{C77}},
  \bibinfo{pages}{025803} (\bibinfo{year}{2008}{\natexlab{a}}).

\bibitem[{\citenamefont{Dexheimer
  et~al.}(2008{\natexlab{b}})\citenamefont{Dexheimer, Schramm, and
  Stoecker}}]{Dexheimer:2008aj}
\bibinfo{author}{\bibfnamefont{V.}~\bibnamefont{Dexheimer}},
  \bibinfo{author}{\bibfnamefont{S.}~\bibnamefont{Schramm}}, \bibnamefont{and}
  \bibinfo{author}{\bibfnamefont{H.}~\bibnamefont{Stoecker}}
  (\bibinfo{year}{2008}{\natexlab{b}}), \eprint{arXiv:0801.2523 [astro-ph]}.

\bibitem[{\citenamefont{Dexheimer
  et~al.}(2008{\natexlab{c}})\citenamefont{Dexheimer, Pagliara, Tolos,
  Schaffner-Bielich, and Schramm}}]{Dexheimer:2008cv}
\bibinfo{author}{\bibfnamefont{V.}~\bibnamefont{Dexheimer}},
  \bibinfo{author}{\bibfnamefont{G.}~\bibnamefont{Pagliara}},
  \bibinfo{author}{\bibfnamefont{L.}~\bibnamefont{Tolos}},
  \bibinfo{author}{\bibfnamefont{J.}~\bibnamefont{Schaffner-Bielich}},
  \bibnamefont{and} \bibinfo{author}{\bibfnamefont{S.}~\bibnamefont{Schramm}},
\bibinfo{journal}{Eur. Phys. J.} \textbf{\bibinfo{volume}{A38}},
  \bibinfo{pages}{105} (\bibinfo{year}{2008}).

\bibitem[{\citenamefont{Mishustin et~al.}(1993)\citenamefont{Mishustin,
  Bondorf, and Rho}}]{Mishustin:1993ub}
\bibinfo{author}{\bibfnamefont{I.}~\bibnamefont{Mishustin}},
  \bibinfo{author}{\bibfnamefont{J.}~\bibnamefont{Bondorf}}, \bibnamefont{and}
  \bibinfo{author}{\bibfnamefont{M.}~\bibnamefont{Rho}},
  \bibinfo{journal}{Nucl. Phys.} \textbf{\bibinfo{volume}{A555}},
  \bibinfo{pages}{215} (\bibinfo{year}{1993}).

\bibitem[{\citenamefont{Furnstahl et~al.}(1995)\citenamefont{Furnstahl, Tang,
  and Serot}}]{Furnstahl:1995by}
\bibinfo{author}{\bibfnamefont{R.~J.} \bibnamefont{Furnstahl}},
  \bibinfo{author}{\bibfnamefont{H.-B.} \bibnamefont{Tang}}, \bibnamefont{and}
  \bibinfo{author}{\bibfnamefont{B.~D.} \bibnamefont{Serot}},
  \bibinfo{journal}{Phys. Rev.} \textbf{\bibinfo{volume}{C52}},
  \bibinfo{pages}{1368} (\bibinfo{year}{1995}).

\bibitem[{\citenamefont{Papazoglou et~al.}(1998)\citenamefont{Papazoglou,
  Schramm, Schaffner-Bielich, Stoecker, and Greiner}}]{Papazoglou:1997uw}
\bibinfo{author}{\bibfnamefont{P.}~\bibnamefont{Papazoglou}},
  \bibinfo{author}{\bibfnamefont{S.}~\bibnamefont{Schramm}},
  \bibinfo{author}{\bibfnamefont{J.}~\bibnamefont{Schaffner-Bielich}},
  \bibinfo{author}{\bibfnamefont{H.}~\bibnamefont{Stoecker}}, \bibnamefont{and}
  \bibinfo{author}{\bibfnamefont{W.}~\bibnamefont{Greiner}},
  \bibinfo{journal}{Phys. Rev.} \textbf{\bibinfo{volume}{C57}},
  \bibinfo{pages}{2576} (\bibinfo{year}{1998}).

\bibitem[{\citenamefont{Papazoglou et~al.}(1999)}]{Papazoglou:1998vr}
\bibinfo{author}{\bibfnamefont{P.}~\bibnamefont{Papazoglou}}
  \bibnamefont{et~al.}, \bibinfo{journal}{Phys. Rev.}
  \textbf{\bibinfo{volume}{C59}}, \bibinfo{pages}{411} (\bibinfo{year}{1999}).

\bibitem[{\citenamefont{Wang et~al.}(2004)\citenamefont{Wang, Lyubovitskij,
  Gutsche, and Faessler}}]{Wang:2003cn}
\bibinfo{author}{\bibfnamefont{P.}~\bibnamefont{Wang}},
  \bibinfo{author}{\bibfnamefont{V.~E.} \bibnamefont{Lyubovitskij}},
  \bibinfo{author}{\bibfnamefont{T.}~\bibnamefont{Gutsche}}, \bibnamefont{and}
  \bibinfo{author}{\bibfnamefont{A.}~\bibnamefont{Faessler}},
  \bibinfo{journal}{Phys. Rev.} \textbf{\bibinfo{volume}{C70}},
  \bibinfo{pages}{015202} (\bibinfo{year}{2004}).

\bibitem[{\citenamefont{Dexheimer
  et~al.}(2008{\natexlab{d}})\citenamefont{Dexheimer, Schramm, and
  Stoecker}}]{Dexheimer:2007df}
\bibinfo{author}{\bibfnamefont{V.}~\bibnamefont{Dexheimer}},
  \bibinfo{author}{\bibfnamefont{S.}~\bibnamefont{Schramm}}, \bibnamefont{and}
  \bibinfo{author}{\bibfnamefont{H.}~\bibnamefont{Stoecker}},
  \bibinfo{journal}{J. Phys.} \textbf{\bibinfo{volume}{G35}},
  \bibinfo{pages}{014060} (\bibinfo{year}{2008}{\natexlab{d}}).

\bibitem[{\citenamefont{Dexheimer and Schramm}(2008)}]{Dexheimer:2008ax}
\bibinfo{author}{\bibfnamefont{V.}~\bibnamefont{Dexheimer}} \bibnamefont{and}
  \bibinfo{author}{\bibfnamefont{S.}~\bibnamefont{Schramm}}
  (\bibinfo{year}{2008}), \eprint{arXiv:0802.1999 [astro-ph]}.

\bibitem[{\citenamefont{Bonanno et~al.}(2007)\citenamefont{Bonanno, Drago, and
  Lavagno}}]{Bonanno:2007kh}
\bibinfo{author}{\bibfnamefont{L.}~\bibnamefont{Bonanno}},
  \bibinfo{author}{\bibfnamefont{A.}~\bibnamefont{Drago}}, \bibnamefont{and}
  \bibinfo{author}{\bibfnamefont{A.}~\bibnamefont{Lavagno}},
  \bibinfo{journal}{Phys. Rev. Lett.} \textbf{\bibinfo{volume}{99}},
  \bibinfo{pages}{242301} (\bibinfo{year}{2007}).

\bibitem[{\citenamefont{Drago et~al.}(2004)\citenamefont{Drago, Gibilisco, and
  Ratti}}]{Drago:2001gd}
\bibinfo{author}{\bibfnamefont{A.}~\bibnamefont{Drago}},
  \bibinfo{author}{\bibfnamefont{M.}~\bibnamefont{Gibilisco}},
  \bibnamefont{and} \bibinfo{author}{\bibfnamefont{C.}~\bibnamefont{Ratti}},
  \bibinfo{journal}{Nucl. Phys.} \textbf{\bibinfo{volume}{A742}},
  \bibinfo{pages}{165} (\bibinfo{year}{2004}).

\bibitem[{\citenamefont{Schechter}(1980)}]{Schechter:1980ak}
\bibinfo{author}{\bibfnamefont{J.}~\bibnamefont{Schechter}},
  \bibinfo{journal}{Phys. Rev.} \textbf{\bibinfo{volume}{D21}},
  \bibinfo{pages}{3393} (\bibinfo{year}{1980}).

\bibitem[{\citenamefont{Migdal and Shifman}(1982)}]{Migdal:1982jp}
\bibinfo{author}{\bibfnamefont{A.~A.} \bibnamefont{Migdal}} \bibnamefont{and}
  \bibinfo{author}{\bibfnamefont{M.~A.} \bibnamefont{Shifman}},
  \bibinfo{journal}{Phys. Lett.} \textbf{\bibinfo{volume}{B114}},
  \bibinfo{pages}{445} (\bibinfo{year}{1982}).

\bibitem[{\citenamefont{Heide et~al.}(1992)\citenamefont{Heide, Rudaz, and
  Ellis}}]{Heide:1992tk}
\bibinfo{author}{\bibfnamefont{E.~K.} \bibnamefont{Heide}},
  \bibinfo{author}{\bibfnamefont{S.}~\bibnamefont{Rudaz}}, \bibnamefont{and}
  \bibinfo{author}{\bibfnamefont{P.~J.} \bibnamefont{Ellis}},
  \bibinfo{journal}{Phys. Lett.} \textbf{\bibinfo{volume}{B293}},
  \bibinfo{pages}{259} (\bibinfo{year}{1992}).

\bibitem[{\citenamefont{Ellis et~al.}(1992)\citenamefont{Ellis, Heide, and
  Rudaz}}]{Ellis:1992ey}
\bibinfo{author}{\bibfnamefont{P.~J.} \bibnamefont{Ellis}},
  \bibinfo{author}{\bibfnamefont{E.~K.} \bibnamefont{Heide}}, \bibnamefont{and}
  \bibinfo{author}{\bibfnamefont{S.}~\bibnamefont{Rudaz}},
  \bibinfo{journal}{Phys. Lett.} \textbf{\bibinfo{volume}{B282}},
  \bibinfo{pages}{271} (\bibinfo{year}{1992}).

\bibitem[{\citenamefont{Brown and Rho}(1991)}]{Brown:1991kk}
\bibinfo{author}{\bibfnamefont{G.~E.} \bibnamefont{Brown}} \bibnamefont{and}
  \bibinfo{author}{\bibfnamefont{M.}~\bibnamefont{Rho}},
  \bibinfo{journal}{Phys. Rev. Lett.} \textbf{\bibinfo{volume}{66}},
  \bibinfo{pages}{2720} (\bibinfo{year}{1991}).

\bibitem[{\citenamefont{Mocsy et~al.}(2004)\citenamefont{Mocsy, Mishustin, and
  Ellis}}]{Mocsy:2004ab}
\bibinfo{author}{\bibfnamefont{A.}~\bibnamefont{Mocsy}},
  \bibinfo{author}{\bibfnamefont{I.~N.} \bibnamefont{Mishustin}},
  \bibnamefont{and} \bibinfo{author}{\bibfnamefont{P.~J.} \bibnamefont{Ellis}},
  \bibinfo{journal}{Phys. Rev.} \textbf{\bibinfo{volume}{C70}},
  \bibinfo{pages}{015204} (\bibinfo{year}{2004}).

\bibitem[{\citenamefont{Kapusta and Gale}(2006)}]{Kapusta:2006pm}
\bibinfo{author}{\bibfnamefont{J.~I.} \bibnamefont{Kapusta}} \bibnamefont{and}
  \bibinfo{author}{\bibfnamefont{C.}~\bibnamefont{Gale}},
  \emph{\bibinfo{title}{{Finite-temperature field theory: Principles and
  applications}}} (\bibinfo{year}{2006}), \bibinfo{note}{{Cambridge University
  Press}}.

\bibitem[{\citenamefont{Cheng et~al.}(2007)}]{Cheng:2006aj}
\bibinfo{author}{\bibfnamefont{M.}~\bibnamefont{Cheng}} \bibnamefont{et~al.},
  \bibinfo{journal}{Phys. Rev.} \textbf{\bibinfo{volume}{D75}},
  \bibinfo{pages}{034506} (\bibinfo{year}{2007}).

\bibitem[{\citenamefont{Cheng et~al.}(2006)}]{Cheng:2006qk}
\bibinfo{author}{\bibfnamefont{M.}~\bibnamefont{Cheng}} \bibnamefont{et~al.},
  \bibinfo{journal}{Phys. Rev.} \textbf{\bibinfo{volume}{D74}},
  \bibinfo{pages}{054507} (\bibinfo{year}{2006}).

\bibitem[{\citenamefont{Blum et~al.}(1996)\citenamefont{Blum, Hetrick, and
  Toussaint}}]{Blum:1995cb}
\bibinfo{author}{\bibfnamefont{T.~C.} \bibnamefont{Blum}},
  \bibinfo{author}{\bibfnamefont{J.~E.} \bibnamefont{Hetrick}},
  \bibnamefont{and}
  \bibinfo{author}{\bibfnamefont{D.}~\bibnamefont{Toussaint}},
  \bibinfo{journal}{Phys. Rev. Lett.} \textbf{\bibinfo{volume}{76}},
  \bibinfo{pages}{1019} (\bibinfo{year}{1996}).

\bibitem[{\citenamefont{Arsene et~al.}(2007)}]{Arsene:2006vf}
\bibinfo{author}{\bibfnamefont{I.~C.} \bibnamefont{Arsene}}
  \bibnamefont{et~al.}, \bibinfo{journal}{Phys. Rev.}
  \textbf{\bibinfo{volume}{C75}}, \bibinfo{pages}{034902}
  (\bibinfo{year}{2007}).

\bibitem[{\citenamefont{Glendenning and Moszkowski}(1991)}]{Glendenning:1991es}
\bibinfo{author}{\bibfnamefont{N.~K.} \bibnamefont{Glendenning}}
  \bibnamefont{and} \bibinfo{author}{\bibfnamefont{S.~A.}
  \bibnamefont{Moszkowski}}, \bibinfo{journal}{Phys. Rev. Lett.}
  \textbf{\bibinfo{volume}{67}}, \bibinfo{pages}{2414} (\bibinfo{year}{1991}).

\bibitem[{\citenamefont{Russkikh and Ivanov}(2006)}]{Russkikh:2006ae}
\bibinfo{author}{\bibfnamefont{V.~N.} \bibnamefont{Russkikh}} \bibnamefont{and}
  \bibinfo{author}{\bibfnamefont{Y.~B.} \bibnamefont{Ivanov}},
  \bibinfo{journal}{Phys. Rev.} \textbf{\bibinfo{volume}{C74}},
  \bibinfo{pages}{034904} (\bibinfo{year}{2006}).

\bibitem[{\citenamefont{Geissel et~al.}(2002)}]{Geissel:2001px}
\bibinfo{author}{\bibfnamefont{H.}~\bibnamefont{Geissel}} \bibnamefont{et~al.},
  \bibinfo{journal}{AIP Conf. Proc.} \textbf{\bibinfo{volume}{619}},
  \bibinfo{pages}{749} (\bibinfo{year}{2002}).

\bibitem[{\citenamefont{Friedman and Gal}(1998)}]{Friedman:1998ed}
\bibinfo{author}{\bibfnamefont{E.}~\bibnamefont{Friedman}} \bibnamefont{and}
  \bibinfo{author}{\bibfnamefont{A.}~\bibnamefont{Gal}},
  \bibinfo{journal}{Phys. Lett.} \textbf{\bibinfo{volume}{B432}},
  \bibinfo{pages}{235} (\bibinfo{year}{1998}).

\bibitem[{\citenamefont{Trnka et~al.}(2005)}]{Trnka:2005ey}
\bibinfo{author}{\bibfnamefont{D.}~\bibnamefont{Trnka}} \bibnamefont{et~al.}
  (\bibinfo{collaboration}{CBELSA/TAPS}), \bibinfo{journal}{Phys. Rev. Lett.}
  \textbf{\bibinfo{volume}{94}}, \bibinfo{pages}{192303}
  (\bibinfo{year}{2005}).

\bibitem[{\citenamefont{Muto et~al.}(2007)}]{Muto:2005za}
\bibinfo{author}{\bibfnamefont{R.}~\bibnamefont{Muto}} \bibnamefont{et~al.}
  (\bibinfo{collaboration}{KEK-PS-E325}), \bibinfo{journal}{Phys. Rev. Lett.}
  \textbf{\bibinfo{volume}{98}}, \bibinfo{pages}{042501}
  (\bibinfo{year}{2007}).

\bibitem[{\citenamefont{Metag}(2007)}]{Metag:2007zz}
\bibinfo{author}{\bibfnamefont{V.}~\bibnamefont{Metag}}, \bibinfo{journal}{J.
  Phys.} \textbf{\bibinfo{volume}{G34}}, \bibinfo{pages}{S397}
  (\bibinfo{year}{2007}).

\bibitem[{\citenamefont{Nasseripour et~al.}(2007)}]{:2007mga}
\bibinfo{author}{\bibfnamefont{R.}~\bibnamefont{Nasseripour}}
  \bibnamefont{et~al.} (\bibinfo{collaboration}{CLAS}), \bibinfo{journal}{Phys.
  Rev. Lett.} \textbf{\bibinfo{volume}{99}}, \bibinfo{pages}{262302}
  (\bibinfo{year}{2007}).

\bibitem[{\citenamefont{Adamova et~al.}(2003)}]{Adamova:2002kf}
\bibinfo{author}{\bibfnamefont{D.}~\bibnamefont{Adamova}} \bibnamefont{et~al.}
  (\bibinfo{collaboration}{CERES/NA45}), \bibinfo{journal}{Phys. Rev. Lett.}
  \textbf{\bibinfo{volume}{91}}, \bibinfo{pages}{042301}
  (\bibinfo{year}{2003}).

\bibitem[{\citenamefont{Arnaldi et~al.}(2006)}]{Arnaldi:2006jq}
\bibinfo{author}{\bibfnamefont{R.}~\bibnamefont{Arnaldi}} \bibnamefont{et~al.}
  (\bibinfo{collaboration}{NA60}), \bibinfo{journal}{Phys. Rev. Lett.}
  \textbf{\bibinfo{volume}{96}}, \bibinfo{pages}{162302}
  (\bibinfo{year}{2006}).

\bibitem[{\citenamefont{van Hees and Rapp}(2006)}]{vanHees:2006iv}
\bibinfo{author}{\bibfnamefont{H.}~\bibnamefont{van Hees}} \bibnamefont{and}
  \bibinfo{author}{\bibfnamefont{R.}~\bibnamefont{Rapp}}
  (\bibinfo{year}{2006}), \eprint{hep-ph/0604269}.


\bibitem[{\citenamefont{Woosley and Janka}(2005)}]{Woosley:2006ie}
\bibinfo{author}{\bibfnamefont{S.}~\bibnamefont{Woosley}} \bibnamefont{and}
  \bibinfo{author}{\bibfnamefont{T.}~\bibnamefont{Janka}},
  \bibinfo{journal}{Nature Physics} \textbf{\bibinfo{volume}{1}},
  \bibinfo{pages}{147} (\bibinfo{year}{2005}).

\bibitem[{\citenamefont{Swesty et~al.}(1994)\citenamefont{Swesty, Lattimer, and
  Myra}}]{SLM}
\bibinfo{author}{\bibfnamefont{F.~D.} \bibnamefont{Swesty}},
  \bibinfo{author}{\bibfnamefont{J.~M.} \bibnamefont{Lattimer}},
  \bibnamefont{and} \bibinfo{author}{\bibfnamefont{E.~S.} \bibnamefont{Myra}},
  \bibinfo{journal}{Apj} \textbf{\bibinfo{volume}{425}}, \bibinfo{pages}{195}
  (\bibinfo{year}{1994}).

\bibitem[{\citenamefont{Baron et~al.}(1985)\citenamefont{Baron, Cooperstein,
  and Kahana}}]{Baron:1985gg}
\bibinfo{author}{\bibfnamefont{E.}~\bibnamefont{Baron}},
  \bibinfo{author}{\bibfnamefont{J.}~\bibnamefont{Cooperstein}},
  \bibnamefont{and} \bibinfo{author}{\bibfnamefont{S.}~\bibnamefont{Kahana}},
  \bibinfo{journal}{Phys. Rev. Lett.} \textbf{\bibinfo{volume}{55}},
  \bibinfo{pages}{126} (\bibinfo{year}{1985}).

\bibitem[{\citenamefont{Migdal et~al.}(1979)\citenamefont{Migdal, Chernoutsan,
  and Mishustin}}]{Migdal:1979je}
\bibinfo{author}{\bibfnamefont{A.~B.} \bibnamefont{Migdal}},
  \bibinfo{author}{\bibfnamefont{A.~I.} \bibnamefont{Chernoutsan}},
  \bibnamefont{and} \bibinfo{author}{\bibfnamefont{I.~N.}
  \bibnamefont{Mishustin}}, \bibinfo{journal}{Phys. Lett.}
  \textbf{\bibinfo{volume}{B83}}, \bibinfo{pages}{158} (\bibinfo{year}{1979}).

\bibitem[{\citenamefont{Takahara and K.}(1985)}]{Takahara:1985}
\bibinfo{author}{\bibfnamefont{M.}~\bibnamefont{Takahara}} \bibnamefont{and}
  \bibinfo{author}{\bibfnamefont{S.}~\bibnamefont{K.}}, \bibinfo{journal}{Phys.
  Lett.} \textbf{\bibinfo{volume}{B156}}, \bibinfo{pages}{17}
  (\bibinfo{year}{1985}).

\bibitem[{\citenamefont{Gentile et~al.}(1993)\citenamefont{Gentile,
  Aufderheide, Mathews, Swesty, and Fuller}}]{Gentile:1993ma}
\bibinfo{author}{\bibfnamefont{N.~A.} \bibnamefont{Gentile}},
  \bibinfo{author}{\bibfnamefont{M.~B.} \bibnamefont{Aufderheide}},
  \bibinfo{author}{\bibfnamefont{G.~J.} \bibnamefont{Mathews}},
  \bibinfo{author}{\bibfnamefont{F.~D.} \bibnamefont{Swesty}},
  \bibnamefont{and} \bibinfo{author}{\bibfnamefont{G.~M.}
  \bibnamefont{Fuller}}, \bibinfo{journal}{Astrophys. J.}
  \textbf{\bibinfo{volume}{414}}, \bibinfo{pages}{701} (\bibinfo{year}{1993}).

\bibitem[{\citenamefont{Drago}(1999)}]{Drago:1998qu}
\bibinfo{author}{\bibfnamefont{A.}~\bibnamefont{Drago}},
  \bibinfo{journal}{Nucl. Phys.} \textbf{\bibinfo{volume}{A661}},
  \bibinfo{pages}{633} (\bibinfo{year}{1999}).

\bibitem[{\citenamefont{Lattimer and Swesty}(1991)}]{Lattimer:1991nc}
\bibinfo{author}{\bibfnamefont{J.~M.} \bibnamefont{Lattimer}} \bibnamefont{and}
  \bibinfo{author}{\bibfnamefont{F.~D.} \bibnamefont{Swesty}},
  \bibinfo{journal}{Nucl. Phys.} \textbf{\bibinfo{volume}{A535}},
  \bibinfo{pages}{331} (\bibinfo{year}{1991}).


\bibitem[{\citenamefont{Klahn et~al.}(2006)}]{Klahn:2006ir}
\bibinfo{author}{\bibfnamefont{T.}~\bibnamefont{Klahn}} \bibnamefont{et~al.},
  \bibinfo{journal}{Phys. Rev.} \textbf{\bibinfo{volume}{C74}},
  \bibinfo{pages}{035802} (\bibinfo{year}{2006}).

\bibitem[{\citenamefont{Janka}()}]{janka}
\bibinfo{author}{\bibfnamefont{T.}~\bibnamefont{Janka}}, \bibinfo{note}{private
  communication}.

\bibitem[{\citenamefont{Gaitanos et~al.}(2004)}]{Gaitanos:2003zg}
\bibinfo{author}{\bibfnamefont{T.}~\bibnamefont{Gaitanos}}
  \bibnamefont{et~al.}, \bibinfo{journal}{Nucl. Phys.}
  \textbf{\bibinfo{volume}{A732}}, \bibinfo{pages}{24} (\bibinfo{year}{2004}).

\bibitem[{\citenamefont{Liu et~al.}(2002)\citenamefont{Liu, Greco, Baran,
  Colonna, and Di~Toro}}]{Liu:2001iz}
\bibinfo{author}{\bibfnamefont{B.}~\bibnamefont{Liu}},
  \bibinfo{author}{\bibfnamefont{V.}~\bibnamefont{Greco}},
  \bibinfo{author}{\bibfnamefont{V.}~\bibnamefont{Baran}},
  \bibinfo{author}{\bibfnamefont{M.}~\bibnamefont{Colonna}}, \bibnamefont{and}
  \bibinfo{author}{\bibfnamefont{M.}~\bibnamefont{Di~Toro}},
  \bibinfo{journal}{Phys. Rev.} \textbf{\bibinfo{volume}{C65}},
  \bibinfo{pages}{045201} (\bibinfo{year}{2002}).

\bibitem[{\citenamefont{Typel and Wolter}(1999)}]{Typel:1999yq}
\bibinfo{author}{\bibfnamefont{S.}~\bibnamefont{Typel}} \bibnamefont{and}
  \bibinfo{author}{\bibfnamefont{H.~H.} \bibnamefont{Wolter}},
  \bibinfo{journal}{Nucl. Phys.} \textbf{\bibinfo{volume}{A656}},
  \bibinfo{pages}{331} (\bibinfo{year}{1999}).

\bibitem[{\citenamefont{Typel}(2005)}]{Typel:2005ba}
\bibinfo{author}{\bibfnamefont{S.}~\bibnamefont{Typel}},
  \bibinfo{journal}{Phys. Rev.} \textbf{\bibinfo{volume}{C71}},
  \bibinfo{pages}{064301} (\bibinfo{year}{2005}).

\bibitem[{\citenamefont{van Dalen et~al.}(2004)\citenamefont{van Dalen, Fuchs,
  and Faessler}}]{vanDalen:2004pn}
\bibinfo{author}{\bibfnamefont{E.~N.~E.} \bibnamefont{van Dalen}},
  \bibinfo{author}{\bibfnamefont{C.}~\bibnamefont{Fuchs}}, \bibnamefont{and}
  \bibinfo{author}{\bibfnamefont{A.}~\bibnamefont{Faessler}},
  \bibinfo{journal}{Nucl. Phys.} \textbf{\bibinfo{volume}{A744}},
  \bibinfo{pages}{227} (\bibinfo{year}{2004}).

\bibitem[{\citenamefont{van Dalen et~al.}(2005)\citenamefont{van Dalen, Fuchs,
  and Faessler}}]{vanDalen:2005ns}
\bibinfo{author}{\bibfnamefont{E.~N.~E.} \bibnamefont{van Dalen}},
  \bibinfo{author}{\bibfnamefont{C.}~\bibnamefont{Fuchs}}, \bibnamefont{and}
  \bibinfo{author}{\bibfnamefont{A.}~\bibnamefont{Faessler}},
  \bibinfo{journal}{Phys. Rev. Lett.} \textbf{\bibinfo{volume}{95}},
  \bibinfo{pages}{022302} (\bibinfo{year}{2005}).

\bibitem[{\citenamefont{Gross-Boelting
  et~al.}(1999)\citenamefont{Gross-Boelting, Fuchs, and
  Faessler}}]{GrossBoelting:1998jg}
\bibinfo{author}{\bibfnamefont{T.}~\bibnamefont{Gross-Boelting}},
  \bibinfo{author}{\bibfnamefont{C.}~\bibnamefont{Fuchs}}, \bibnamefont{and}
  \bibinfo{author}{\bibfnamefont{A.}~\bibnamefont{Faessler}},
  \bibinfo{journal}{Nucl. Phys.} \textbf{\bibinfo{volume}{A648}},
  \bibinfo{pages}{105} (\bibinfo{year}{1999}).

\bibitem[{\citenamefont{de~Jong and Lenske}(1998)}]{deJong:1997hr}
\bibinfo{author}{\bibfnamefont{F.}~\bibnamefont{de~Jong}} \bibnamefont{and}
  \bibinfo{author}{\bibfnamefont{H.}~\bibnamefont{Lenske}},
  \bibinfo{journal}{Phys. Rev.} \textbf{\bibinfo{volume}{C58}},
  \bibinfo{pages}{890} (\bibinfo{year}{1998}).

\bibitem[{\citenamefont{Kolomeitsev and
  Voskresensky}(2005)}]{Kolomeitsev:2004ff}
\bibinfo{author}{\bibfnamefont{E.~E.} \bibnamefont{Kolomeitsev}}
  \bibnamefont{and} \bibinfo{author}{\bibfnamefont{D.~N.}
  \bibnamefont{Voskresensky}}, \bibinfo{journal}{Nucl. Phys.}
  \textbf{\bibinfo{volume}{A759}}, \bibinfo{pages}{373} (\bibinfo{year}{2005}).

\bibitem[{\citenamefont{Akmal et~al.}(1998)\citenamefont{Akmal, Pandharipande,
  and Ravenhall}}]{Akmal:1998cf}
\bibinfo{author}{\bibfnamefont{A.}~\bibnamefont{Akmal}},
  \bibinfo{author}{\bibfnamefont{V.~R.} \bibnamefont{Pandharipande}},
  \bibnamefont{and} \bibinfo{author}{\bibfnamefont{D.~G.}
  \bibnamefont{Ravenhall}}, \bibinfo{journal}{Phys. Rev.}
  \textbf{\bibinfo{volume}{C58}}, \bibinfo{pages}{1804} (\bibinfo{year}{1998}).

\bibitem[{\citenamefont{Nice et~al.}(2005)}]{Nice:2005fi}
\bibinfo{author}{\bibfnamefont{D.~J.} \bibnamefont{Nice}} \bibnamefont{et~al.},
  \bibinfo{journal}{Astrophys. J.} \textbf{\bibinfo{volume}{634}},
  \bibinfo{pages}{1242} (\bibinfo{year}{2005}).

\bibitem[{\citenamefont{Haensel et~al.}()\citenamefont{Haensel, Potekhin, and
  Yakovlev}}]{Haensel:2007yy}
\bibinfo{author}{\bibfnamefont{P.}~\bibnamefont{Haensel}},
  \bibinfo{author}{\bibfnamefont{A.~Y.} \bibnamefont{Potekhin}}
  \bibnamefont{and} \bibinfo{author}{\bibfnamefont{D.~G.}
  \bibnamefont{Yakovlev}},
  \emph{\bibinfo{title}{{Neutron stars 1: Equation of state and structure}}} 
(\bibinfo{year}{2007}), \bibinfo{note}{{Springer, New York}}.


\bibitem[{\citenamefont{Muller}(1997)}]{Muller:1997tm}
\bibinfo{author}{\bibfnamefont{H.}~\bibnamefont{Muller}},
  \bibinfo{journal}{Nucl. Phys.} \textbf{\bibinfo{volume}{A618}},
  \bibinfo{pages}{349} (\bibinfo{year}{1997}).

\bibitem[{\citenamefont{McLerran and Pisarski}(2007)}]{McLerran:2007qj}
\bibinfo{author}{\bibfnamefont{L.}~\bibnamefont{McLerran}} \bibnamefont{and}
  \bibinfo{author}{\bibfnamefont{R.~D.} \bibnamefont{Pisarski}},
  \bibinfo{journal}{Nucl. Phys.} \textbf{\bibinfo{volume}{A796}},
  \bibinfo{pages}{83} (\bibinfo{year}{2007}).

\end{thebibliography}
\bibliographystyle{apsrev}

\end{document}